\documentclass[twocol]{ametsocV5}
\usepackage{amsmath,amsfonts,amssymb,bm}
\usepackage{mathptmx}
\usepackage{newtxtext}
\usepackage{newtxmath}

\newcommand{\gd}[1]{{\color{black}{#1}}}

\def\bm{\mathbf{m}}
\def\bJ{\mathbf{J}}
\def\bp{\mathbf{p}}
\def\bk{\mathbf{k}}

\def\mI{\mathcal{I}}

\def\mB{\mathcal{B}}

\def\mP{\mathcal{P}}
\def\mF{\mathcal{F}}

\def\mJ{{\mathcal J}}
\def\inn{{\rm in}}
\def\out{{\rm out}}
\def\h{{\rm h}}
\def\v{{\rm v}}
\def\max{{\rm max}}
\def\min{{\rm min}}
\DeclareMathOperator{\sign}{sign}

\title{On the Origins of the Oceanic Ultraviolet Catastrophe}

\authors{Giovanni Dematteis\correspondingauthor{Giovanni Dematteis, dematg@rpi.edu}$\ ^{1}$, Kurt Polzin$\ ^2$ and Yuri V. Lvov$\ ^1$}
\affiliation{$\ ^1$Department  of Mathematical  Sciences, Rensselaer Polytechnic Institute, Troy, New York \\
$\ ^2$Woods Hole Oceanographic Institution, Woods Hole, Massachusetts}

\abstract{ \gd{We provide a first-principles analysis of the energy fluxes in the oceanic internal wavefield. The resulting formula is remarkably similar to} the renowned
  phenomenological formula for the turbulent dissipation rate in the
  ocean which is known as the Finescale Parameterization. The prediction
  is based on the wave turbulence theory of internal gravity waves and
  on a new methodology devised for the computation of the associated
  energy fluxes. In the standard spectral representation of the wave
  energy density, in the two-dimensional vertical wavenumber --
  frequency ($m-\omega$) domain, the energy fluxes associated with the
  steady state are found to be directed downscale in both coordinates,
  closely matching the Finescale-Parameterization formula in
  functional form and in magnitude.  These energy transfers are
  composed of a `local' and a `scale-separated' contributions; while
  the former is quantified numerically, the latter is dominated by the
  Induced Diffusion process and is amenable to analytical
  treatment. Contrary to previous results indicating an inverse energy
  cascade from high frequency to low, at odds with observations, our
  analysis of all non-zero coefficients of the diffusion tensor
  predicts a direct energy cascade. Moreover, by the same analysis
  fundamental spectra that had been deemed `no-flux' solutions are
  reinstated to the status of `constant-downscale-flux'
  solutions. This is consequential for an understanding of energy
  fluxes, sources and sinks that fits in the observational paradigm of
  the Finescale Parameterization, solving at once two long-standing
  paradoxes that had earned the name of `Oceanic Ultraviolet
  Catastrophe'.}

\begin{document}

\maketitle

\section*{Significance Statement}
The Global Circulation Models cannot resolve the scales of the oceanic internal waves. The Finescale Parameterization of turbulent dissipation, a formula grounded in observations, is the standard tool by which the energy transfers due to internal waves are incorporated in the global models. Here, we provide an interpretation of this parameterization formula building on the first-principles statistical theory describing energy transfers between waves at different scales. Our result is in agreement with the Finescale Parameterization, and points out a large contribution to the energy fluxes due to a type of wave interactions (`local') usually disregarded. Moreover, the theory on which the traditional understanding of the parameterization is mainly built, a `diffusion approximation', is known to be partly in contradiction with observations. We put forward a solution to this problem, visualized by means of `streamlines' that improve the intuition of the direction of the energy cascade.

\section{Introduction}\label{sec:Intro}

The intent of this paper is to provide a theoretical \gd{analysis} of the downscale energy transfers
associated with the `Finescale Parameterization' for internal wave breaking,
\cite{gregg1989scaling,henyey1991scaling,polzin1995finescale}.  While there is some underlying discussion of theoretical constructs in those works, application of those theoretical considerations is incomplete and the model is, in essence, heuristic \citep{polzin2004heuristic,polzin2014finescale}.


The crux of the issue is that there is an essential incompatibility between the internal wave spectrum articulated in \cite{GM72} which is separable in frequency and vertical wavenumber {\it vs} analytic theory summarized in \cite{Muller86}, which is based upon extreme scale separated interactions and emphasizes transfers in vertical wavenumber.  We have summarized this intrinsic incompatibility as the `Oceanic Ultraviolet Catastrophe',\gd{~\cite{polzin2017oceanic}}.\footnote{\gd{The parallel with the Ultraviolet Catastrophe of black body radiation is merely in the fact that an assumption of spectral equipartition (of energy density in frequency space, in one case, and of action density in vertical wavenumber space, in the other) leads to a non-physical result: in classical physics this is the prediction of an infinite energy density, while in oceanographic literature it is (among other things discussed shortly) the prediction that the Garrett and Munk spectrum is associated with an equilibrium state, with no fluxes between different scales.} }

There are two aspects to the Oceanic Ultraviolet Catastrophe.  First, that theoretical scenario depicts a transfer of internal wave energy from large to small vertical scales at constant horizontal wavenumber and consequently from high frequency to low \citep{mccomas1981dynamic}.  With such transfer, a source of internal wave energy at high frequency is required for a stationary balance.  However, a systematic review of the nonlinear transfers and possible energy sources of the oceanic internal wavefield \citep{regional} was not able to identify the required source of energy at high frequency \gd{(see also \cite{le2021variability,whalen2020internal,kunze2017internal,ferrari})}.  Secondly, the Garrett and Munk 1976 (GM76) version of the oceanic spectrum, which was given `universal' status in \cite{munk1981internal}, is not just a stationary state in that \citep{Muller86} theoretical paradigm: Having no gradients of action in vertical wavenumber, GM76 is a {\em no flux} solution of the Fokker-Planck equation, which means {\em no transport} of energy to smaller scales.  Yet that same theory makes a prediction for the spectral power laws of statistically stationary states that are in good agreement with observed oceanic spectra, \cite{polzin2017oceanic} (their figure~37). 

These theoretical issues stand in contrast to the finescale parameterization.  The finescale parameterization originates with \cite{gregg1989scaling} as an empirical statement about the ability of 10 meter
first difference estimates of vertical shear to act as a proxy for the dissipation rate $\epsilon$.  It is distinct from both ray tracing simulations \citep{henyey1986energy} and from formal theory using a
characterization of the scale separated interactions \citep{mccomas1981dynamic}.  In \cite{polzin1995finescale} one finds further data/model comparisons, an attempt to address normalization issues, an accounting for departures from the GM frequency distribution using an argument forwarded in \cite{henyey1991scaling} and, importantly, an attempt to place the discussion in the spectral domain rather than using the 10 meter first difference metric.  In so doing there is an assertion that the energy transfers in horizontal wavenumber keep pace with those in vertical wavenumber such that spectral transports do not project strongly across frequencies.  Up to this point the finescale parameterization is interpreted as a model for the refraction of high frequency waves in near-inertial shear.  It can be dissected into one part high frequency energy, one part near inertial shear variance and one part refraction rate proportional to the high frequency wave aspect ratio.  Apart from concerns about the constant out front, these are the same basic ingredients provided by formal theory for extreme scale separated interations and summarized with a Fokker-Plank (or generalized diffusion) equation \citep{polzin2017oceanic}.  In \cite{polzin2004heuristic} one finds a fundamentally distinct interpretation being articulated, that the same finescale parameterization can be viewed as a closure for local, rather than scale separated, interactions.  This characterization is used to find solutions to a boundary source decay problem in \cite{polzin2004idealized} and these solutions are employed to write a dynamically based mixing recipe for the decay of internal tides in \cite{polzin2009abyssal}.

We address the concerns raised by the 'Oceanic Ultraviolet Catastrophy' with theoretical work undertaken in the last decade.  These include numerical estimates that are underpinned by first principles \citep{regional} which suggest a far more nuanced view: there is an obvious role for interactions that are `local' in nature in addition to those that are `extreme scale separated'.  This provides an interpretation that parallels the two (local vs. extreme scale separated) interpretations of the finescale parameterization.  Moreover, there is a growing appreciation that the assessment of the the Garrett and Munk spectrum as a no-flux solution is incomplete \citep{dematteis_lvov_2021}.  Here we build upon the results of \cite{dematteis_lvov_2021} to analyze the energy fluxes in the oceanic internal wave field and provide a first principle explanation of the finescale parameterization.  

The finescale parameterization is
\begin{equation}\label{eq:FinescaleParameterization}
\begin{aligned}
\mathcal{P}_{{\rm finescale}} = 8& \times 10^{-10} \frac{f}{f_0}
\frac{N^2 {\rm cosh}^{-1}(N/f) }{N_0^2 {\rm cosh}^{-1}(N_0/f_0)} \;\\
&\times\hat{E}^2 \; \frac{3(R_{\omega}+1)}{4R_{\omega}}
\sqrt{\frac{2}{R_{\omega}-1}} {\rm W/kg}\,,
\end{aligned}
\end{equation}
in which $\mathcal{P}$ is the downscale energy transport rate, to be
partitioned between kinetic energy dissipation rate $\epsilon$ and
work done against gravity in a buoyancy flux.  The factor $\hat{E} =
0.1 {\rm cpm}/m_c$ is a length scale metric of the shear spectral
density, with vertical wavenumber $m_c$ defined by a transition in
spectral slope \citep{gargett1981composite} to a wave breaking region:
\begin{equation}
\displaystyle \int_o^{m_c} 2m^2 E_k(m) dm = \frac{2\pi}{10 } {\;\rm m}^{-1}
\end{equation}\label{eq:Amplitude}
with horizontal kinetic energy density $E_k(m)$.  The factor
$R_{\omega}$ is the ratio of the gradient potential energy spectrum to
the gradient horizontal kinetic energy spectrum.  Finally, $f_0$ and
$N_0$ are normalization constants for the Coriolis \gd{frequency $f$} and the buoyancy
frequency \gd{$N$}.  The parameterization uses values corresponding to the
local pendulum day at 32.5 degrees latitude and 3 cph. \gd{Formula~\eqref{eq:FinescaleParameterization} is normalized so that for GM76, for which $R_\omega=3, \hat E=1$, for $f=f_0$ and $N=N_0$ one has $\mathcal{P}_{{\rm finescale}} = 8 \times 10^{-10}\;$W$/$kg.}

\begin{figure*}
\begin{centering}
\includegraphics[width=0.9\linewidth]{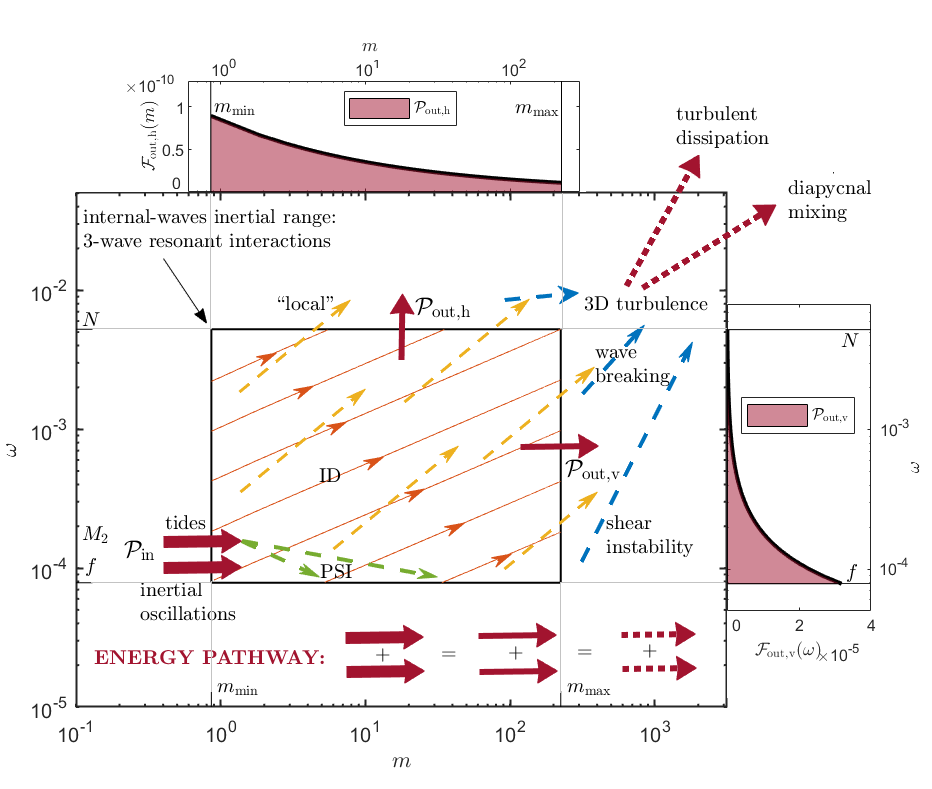}
  \caption{Energy pathways in
    the internal-wave band. Tides and inertial oscillations
    force the near-inertial modes ($\omega\sim f$). The PSI decay
    mechanism is believed to transfer a large amount of energy to
    larger wavenumbers, at frequency $f$, providing a plausible
    physical mechanism for the bottom edge at $\omega=f$ to act as
    an energy source. Resonant interactions between triads of internal
    waves dominate the spectral energy transfers in the `inertial
    range' (inner black box in the figure). These energy fluxes are in
    part diffusive, shown as solid red (analytically obtained)
    streamlines, and in part `local', shown (qualitatively) as yellow
    dashed arrows. The energy transferred to scales smaller than $10$
    m ($m>m_{\max}$) and frequencies above the buoyancy frequency
    ($\omega>N$) are assumed to excite hydrodynamic instabilities that result in a 3D
    turbulent field. The end result is dissipation of turbulent
    kinetic energy and diapycnal mixing due to the work of buoyancy
    fluxes. The power exiting the wavefield, per unit of $\omega$
    at the $m=m_{\max}$ edge, and per unit of $m$ at the $\omega=N$ edge, respectively,
    is represented in the two insets, result of the theoretical
    calculation presented in Sec. \ref{sec:loc-extr}.  In the figure,
    dashed lines represent finite spectral `jumps'.
  }
 \label{fig:0}
\end{centering}
\end{figure*}
For
our calculations, we use a modified version of the GM76 spectrum, which is consistent with the stationary solution
of the kinetic equation \gd{in the scale invariant regime, found in \cite{iwthLPTN}}. Denoting the wavenumber by $\bp=(\bk,\bm)$,
$\bk$ being the two-dimensional horizontal projection and $\bm$ the
vertical projection, whose magnitudes are denoted by $k=|\bk|$ and
$m=|\bm|$, the GM76 model is multiplied by $(k/k_{\ast})^{0.31}$, with $m_\ast=4\pi N/(bN_0)$, 
$k_{\ast} = c m_{\ast} f / N $ (where $c\simeq3$ comes from the constraint that the modified version preserve the same energy level as GM76):
\begin{equation}
e(m,\omega) = \frac{2f}{\pi}\frac{1}{\omega\sqrt{\omega^2-f^2}}
\frac{2m_{\ast}}{\pi} \frac{1}{m_{\ast}^2+m^2} b^2 N N_0 E
\left(\frac{k}{k_{\ast}}\right)^{0.31}\,,
\label{eq:ModifiedGMspectrum}
\end{equation}
\gd{where $E$ is the non-dimensional GM76 energy level. Following the instructions of \cite{polzin1995finescale},} numerical integration of the modified spectrum provides $\hat{E}=2.46$ and $R_{\omega} = 2.48$, the two parameters needed \gd{to calculate Eq.~\eqref{eq:FinescaleParameterization} for the spectrum~\eqref{eq:ModifiedGMspectrum}}, and thus
\begin{equation}\label{eq:Pfinescale}
\mathcal{P}_{{\rm finescale}} \rightarrow 5.9\times 10^{-9} {\rm W/kg}\,.
\end{equation}
The corresponding first principles estimate of \cite{dematteis_lvov_2021}, reviewed and analysed in section 3, is 
\begin{equation}\label{eq:Pdematteis}
\mathcal{P}_{{\rm DL}} \rightarrow 9.0\times 10^{-9} {\rm W/kg}\,.
\end{equation}
These estimates essentially have a $\hat{E}^2 N^2 f$ scaling in
common, although the theoretical estimate will have small exponent
corrections due to the modification to the scaling of the GM76
spectrum (see Eqs.~\eqref{eq:transfer} below). The finescale
parameterization has a logarithmic pre-factor of ${\rm
  cosh}^{-1}(N/f)$ whilst the theoretical estimate leading to Eq.~\eqref{eq:Pdematteis}
contains a power series in $f/N$.  The relative agreement of the first
principles estimate $\mathcal{P}_{{\rm DL}}$ with $\mathcal{P}_{{\rm
    finescale}}$ requires interpretation and discussion.

The first principles analysis provides
us with more than the simple downscale transport rate~\eqref{eq:Pdematteis}. The downscale direction of the energy fluxes,
both in vertical wavenumber {\it and} in frequency \gd{(in agreement with the recent results by \cite{eden2019numerical})}, and a novel
explanation of the {\it no-flux} paradox in the Fokker-Planck
paradigm, will allow us to propose a solution to the Oceanic
Ultraviolet Catastrophe. The estimate springs from the wave turbulence
kinetic equation governing transfers within a spectrum of amplitude
modulated waves, and fits within a general picture schematized in
Fig. \ref{fig:0}, to which we will refer in the rest of the
manuscript.  The energy of the large scale inertial oscillations and
tides (of the order of a cycle per day) is transferred between
interacting internal gravity waves. The mechanism of nonlinear
resonant interaction between internal wave triads is assumed to
dominate the scene in an `inertial range' (in the sense of turbulence,
i.e. a range of scales where no other effects such as forcing or
damping are present) extending down to the buoyancy frequency scales
(several cycles per hour) and, in terms of vertical scales, spanning
from the ocean depth (several kilometers) to the wave breaking scale
(around 10 meters). These resonant transfers result in a downscale
energy flux both in frequency and vertical wavenumber.  Part of this
transfer can be approximated as a pointwise flux due to the
scale-separated Induced-diffusion process. The streamlines of the
diffusive part of the flux (analytically obtained, see
Sec. \ref{sec:ID}) are represented as solid red lines in
Fig. \ref{fig:0}. The contribution to the flux by `local'
interactions, which is given by finite `jumps' between separate
points in Fourier space, is represented (qualitatively\gd{, in the schematic of Fig.~\ref{fig:0}}) by yellow
dashed arrows.
Lastly, turbulent instabilities at smaller scales mark the end of the
cascade of energy, which finally goes into the work of buoyancy fluxes
against gravity, generating diapycnal mixing, and into dissipated
turbulent kinetic energy.

The stationary state identified in \cite{iwthLPTN} is supported by a mixture of both
`local' and `scale separated' interactions.  In section
\ref{sec:loc-extr} we consider both types of interactions and separate
the (non-rotating) transports (\ref{eq:Pdematteis}) into the
respective fluxes, in quantitative agreement with the finescale
parameterization. We \gd{locate} the separation between the
two types of interactions and we show that the `scale separated' part
reduces correctly to the diffusive prediction.

We then overview the internal wave kinetic equation and discuss questions of stationary
states, inertial ranges, and convergence of the associated integrals in section
\ref{sec:WKE}.  In section \ref{sec:ID} we revisit the energy flux
theory of the Fokker-Planck equation in the Induced-Diffusion limit. We analyze the
relation between horizontal and vertical wavenumber fluxes and discuss
how these transfers project onto the frequency domain, crucially requiring an energy source at low frequency.  In
section \ref{sec:TheEnd} we summarize our results and suggest
a way out of the paradox referred to as the Oceanic Ultraviolet
Catastrophe.

\section{Local {\it vs} scale-separated contributions to the energy fluxes}\label{sec:loc-extr}

We consider the internal wave kinetic equation  in the scale-invariant regime, consisting of neglecting
the effects of the Coriolis force. An idealized stratified ocean
without spatial inhomogeneities is assumed, in the isopycnal
representation consisting of the use of the mass density $\rho$ as
vertical coordinate in place of the water depth $z$. Thus, $\bk$ is in
units of meters$^{-1}$ and $\bm$ is in units of meters$^3/$kg. The
problem is further simplified by considering a constant stratification
profile and an isotropic wave field in the horizontal directions.  The
non-rotating dispersion relation of internal waves reduces to
\begin{equation}\label{eq:disp}
    \omega=\gamma k/m\,, \qquad \text{with}\qquad \gamma=g/(\rho_0 N)\,,
    \end{equation}
where $g$ is the acceleration of gravity and $\rho_0$ is the reference
mass density.  The statistical quantities characterizing this
homogeneous, horizontally isotropic wavefield are: the 3D spectral
action density $n(\bp)$, the 2D spectral action density $n(k,m)=4\pi k
n(\bp)$, and the 2D spectral energy density $e(k,m)=\omega n(k,m)$. At
convenience, one can switch from the $k-m$ space representation to the
$\omega-m$ space representation. The change of coordinates is simply
defined by the dispersion relation \eqref{eq:disp}:
$n(\omega,m)=n(k,m)\left(\frac{\partial \omega}{\partial
  k}\right)^{-1}$, $e(\omega,m)=e(k,m)\left(\frac{\partial
  \omega}{\partial k}\right)^{-1}$ -- note that the latter quantity
has been used in the introduction in
Eq. \eqref{eq:ModifiedGMspectrum}.

We consider the stationary solution \eqref{eq:ModifiedGMspectrum}, which translates into a 3D action spectral density
of the form
\begin{equation}\label{eq:powlaw}
    n(\bp)=Ak^{-a}m^{-b}\,,\quad a=3.69,\;\;b=0\,.
\end{equation}
The internal wave kinetic equation expresses the time evolution of the 3D action due to 3-wave nonlinear resonant interactions, in a way that will be detailed in Sec. \ref{sec:WKE}.
The equation can be written as
\begin{equation}\label{eq:kineq_decomp}
   \frac{\partial n_\bp}{\partial t} = \mI^{({\rm loc})} + \mI^{({\rm sep})}\,,
\end{equation}
according to the classical decomposition put forward by
\cite{McComas1977} into `local' and `scale separated' interactions. In
particular, the latter kind of interactions is dominated, in a
spectrum close to equilibrium, by the Induced Diffusion (ID) process,
which allows one to simplify its contribution to an actual diffusion
such that
\begin{equation}\label{eq:IDrhs}
   \mI^{({\rm sep})} \simeq \frac{\partial }{\partial p_i} \left( a_{ij}\frac{\partial n_\bp}{\partial p_j} \right)\,,
\end{equation}
where $a_{ij}$ is the diffusion tensor and $i,j=1,2,3$ denote the three components of the wavevector $\bp$.

Let us consider the inner box delimited by a solid black line in
Fig. \ref{fig:0} and refer to it as the `inertial range', denoted by
$\mB$, in $k-m$ space rather than in $m-\omega$ simply for ease of
calculation. Since there are no sources or sinks of energy inside
$\mB$, one can write the energy conservation equation in integral form
for $\mB$, as
\begin{equation}\label{eq:int_cons}
\begin{aligned}
	&\frac{d}{d t}\int_\mB e(k,m)\; d k\, d m + \mP_{\inn} +
  \mP_{\out}=0\,,\\ &\mP_{\inn}=\int_{\partial\mB_{\inn}}\mF(s)\,ds\,,
  \quad \mP_{\out}=\int_{\partial\mB_{\out}}\mF(s)\,ds\,,
	\end{aligned}
\end{equation}
\begin{figure*}
\begin{centering}
\includegraphics[width=\linewidth]{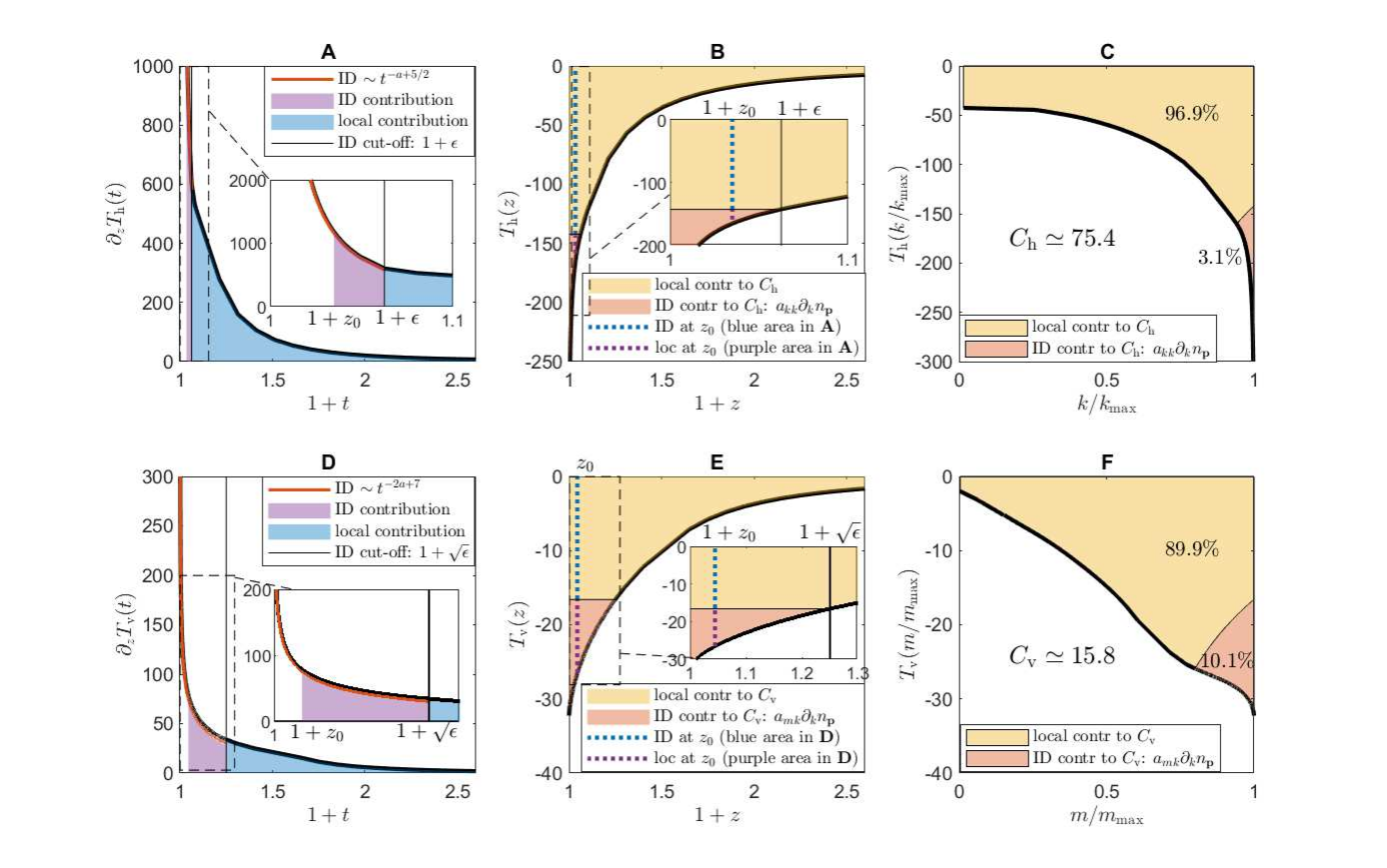}
  \caption{Construction of the transfer integrals
    $C_{\rm h}$ (upper panels) and $C_{\rm v}$ (lower panels) relative
    to the energy flux at the upper and right edges of the inertial
    box in Fig. \ref{fig:0}, respectively. The contributions in panels
    {\bf A} and {\bf D} are computed numerically, except for the ID
    singularity that is computed analytically. Integration of the
    functions in {\bf A} and {\bf D} gives respectively the
    (nondimensional) transfer integrands in {\bf B} and {\bf E},
    respectively, where the red area denotes the contribution that
    comes from the scale-separated region, dominated by ID. This red
    area, representing the diffusive part of the energy fluxes, is an
    explicit function of the diffusion coefficients, as shown in the
    legend. Panels {\bf C} and {\bf F} are a remapping of {\bf B} and
    {\bf E}, respectively, upon suitable change of coordinates; in
    this physically more intuitive representation, the contribution
    from the right corner at $1$ represents energy transferred across
    the boundary from a neighborhood of the boundary
    itself. Contributions from the left side of the plot in panel {\bf C} and {\bf F}, instead, are due to
    large jumps in spectral space. Again, the red area represents the part of
    the contribution due to ID scattering.}
 \label{fig:1}
\end{centering}
\end{figure*}
where $s$ is a parameterization of the boundary
$\partial\mB=\partial\mB_{\inn}\cup\partial\mB_{\out}$, with
$\partial\mB_{\inn}$ the part of the boundary where $\mF>0$ (energy
entering $\mB$) and $\partial\mB_{\out}$ the part of the boundary
where $\mF<0$ (energy exiting $\mB$). $\mF$ is the power per unit of
$s$ flowing across the boundary, so that $\mP_{\inn}>0$ and
$\mP_{\out}<0$ represent the total power going in and out of $\mB$,
respectively, due to 3-wave nonlinear interactions.

The fluxes in
Eq.~\eqref{eq:int_cons} can be computed directly from the collision
integral, i.e. the r.h.s. of the wave kinetic
equation~\eqref{eq:kineq_decomp}. \gd{The details on the theory and numerics of the method can be found in \cite{dematteis_lvov_2021}, section 5.3. In addition, the ({\it Matlab}) numerical codes can be found as Supplementary Material,~\cite{JPOsupplemental}}. An accurate (numerical) counting of
resonances transferring energy past the $\partial \mB_{\out}$ part of
the boundary leads to the following formulae for the
horizontal (in the $k$ direction, across the upper edge at
$\omega_{\max}$ in Fig.~\ref{fig:0}) and vertical (in the $m$
direction, across the right edge at $m_{\max}$ in Fig.~\ref{fig:0})
outgoing fluxes, respectively:

\begin{equation}\label{eq:transfer}
\begin{aligned}
    &\mP_{\out,\h} = \int_{m_{\min}}^{m_{\max}} d m\, \mF_{\out,h}(m)\,,\\
    &\mP_{\out,\v} = \int_{\frac{f}{\gamma}m_{\max}}^{\frac{N}{\gamma} m_{\max}} d k\, \mF_{\out,v}(k)\,,\\
	&\left(\begin{array}{c}\mF_{\out,\h}(m)\\\mF_{\out,\v}(k)\end{array} \right) = 4\pi\frac{N^2}{g}(V_0A)^2 \left(\begin{array}{c}k_{\max}^{7-2a} C_\h\\ k^{6-2a}m_{\max}C_\v \end{array}\right)\,,\\
   &C_\h = \int_{\frac{f}{N}}^1 ds\, T_\h(s)\,,\quad  C_\v = \int_{\frac{m_{\min}}{m_\max}}^1 ds\, T_\v(s)\,,
   \end{aligned}
\end{equation}
where \gd{$A$ is the prefactor of the action spectrum~\eqref{eq:powlaw}, $V_0$ is a dimensional prefactor of the matrix elements of the kinetic equation (see Eq.~\eqref{eq:powtot} for the explicit expressions of $A$ and $V_0$) and} $T_\h$ and $T_\v$, plotted in Fig. \ref{fig:1} \gd{\cite{JPOsupplemental}}, are \gd{non-dimensional} transfer
integrands quantifying how far from the boundary the outgoing energy
is coming from. \gd{In Fig. \ref{fig:1},} the upper panels are for horizontal energy transport
across the boundary at $\omega=N$, which via the dispersion relation
maps to $k_{\max}(m) = \frac{m N}{\gamma}$. The lower panels are for
vertical energy transport across the boundary at $k=k_{\max}$. Panels
{\bf A} and {\bf B} depict the procedure of double integration of the
collision integrand that leads to the horizontal flux. Panel {\bf C}
is a remapping of panel {\bf B} and quantifies the energy transferred
beyond $k=k_{\max}$ from different distances from the boundary. The
analogous calculation is represented in the lower panels \gd{{\bf D}-{\bf F}} for the
vertical transport beyond \gd{the $m=m_{\max}$ edge}. Note that we use the variable
in brackets to denote different functions $T_{\h/\v}(s), s\in[0,1]$ \gd{(panels {\bf C} and~{\bf F})
and $T_{\h/\v}(s'), s'\in[0,\infty]$ (panels \gd{{\bf B} and {\bf E}})}. As one might expect, the
contribution \gd{to the outgoing flux coming} from the immediate vicinity of the boundary \gd{($k/k_\max=1$ in panel {\bf C} and $m/m_\max=1$ in panel {\bf F})} is due to the
ID process: a wave close to the boundary is scattered right across it,
while at the same time absorbing a much smaller wave number that
`induces' the scattering. This part of the contribution is represented
as the red area in the panels {\bf C} and {\bf F}, for the horizontal
and the vertical fluxes, respectively. These red areas have an
analytical expression in terms of the coefficients of the diffusion
tensor, respectively of $a_{kk}\frac{\partial n_\bp}{\partial k}$ and
$a_{mk}\frac{\partial n_\bp}{\partial k}$. The terms involving
$\frac{\partial n_\bp}{\partial m}$ are identically zero since the
analyzed spectrum, Eq.~\eqref{eq:powlaw} does not depend on
$m$. Looking at panels {\bf A} and {\bf D}, one finds that these
contributions come from the integration of integrable singularities
given by the ID asymptotics, which were quantified numerically in \cite{dematteis_lvov_2021}. We provide the following newly obtained
analytical result:
\begin{equation}\label{eq:coeff_diff}
   a_{kk} \simeq 8\pi \times 5.8 \epsilon^{\frac92-a} k^{6-a}m\,,\quad  a_{mk} \simeq 8\pi \times 4.0 \epsilon^{\frac92-a}k^{5-a}m^2\,,
\end{equation}
that will be explained in more detail in Sec.~\ref{sec:ID}. The flux due to ID (red areas of Fig. \ref{fig:1}, panels {\bf C} and {\bf F}) is therefore given explicitely by (minus) the term in round brackets in Eq.~\eqref{eq:IDrhs}.
Supplementing the numerical integration of the local interactions by the
exact analytical integration of the ID singularities, one obtains:
\begin{equation}
    C_\h\simeq -8\pi\times\,75.4,\qquad C_\v\simeq -8\pi\times\,15.8\,.\footnote{\gd{Here and in the following, a factor of $8\pi$ is kept separated from the result of numerical integration of Eq.~\eqref{eq:kineq} below.}}
\end{equation}
\gd{We point the reader to the Appendix for details of the calculation}. The computation is performed in
horizontal wavenumber variables (vertical wavenumbers are bounded to
the horizontal via the resonant conditions, see Sec. \ref{sec:WKE}), and we estimate that
the power series upon which the ID leading-order approximation is
based holds for points with $k/k_{\max}>(1+\epsilon)^{-1},\,
\epsilon\simeq1/16$. This is what delimits the red ID region in
Fig.~\ref{fig:1}{\bf C}. The ID asymptotics establish that the
scattering of point $(k,m)$ via ID interaction results into a point
$(k(1+O(\epsilon)), m(1+O(\sqrt{\epsilon})))$, for $\epsilon \ll1$, as
represented in Fig.~\ref{fig:1}{\bf D}. With the due changes of
variables, this implies that the ID region for vertical transport is
given by the red area in Fig.~\ref{fig:1}{\bf F}. Therefore, we can
now interpret both transfer integrals $C_\h$ and $C_\v$ as given by \gd{a {\it scale-separated} contribution (dominated by the ID process)} and by a {\it local} contribution. In particular, the
horizontal transport contribution is about $96.9\%$ local, $3.1\%$ ID,
and the vertical transport contribution is about $89.9\%$ local,
$10.1\%$ ID.

Taking into account all interactions and the explicit expressions of
$V_0$ and $A$,
Eq. \eqref{eq:transfer} can be rewritten as
\begin{equation}\label{eq:powtot}
    \begin{aligned}
    &\mP_{\out,\rm h} =
  \frac{\Gamma C_\h}{1-\nu} \left[1-\left(\frac{\ell}{2b}\right)^{1-\nu}\right] f^{1+\nu} N  E^2 \,,\\
  &\mP_{\out,\rm v} = \frac{\Gamma C_\v}{\nu} 
  \left[1-\left(\frac{f}{N}\right)^{\nu}\right] f N^{1+\nu}  E^2  \,,\\
  &\Gamma = \frac{1}{\pi^3}\left(\frac{2 \ell }{b}\right)^\nu \frac{ N_0^{1-\nu} b^3 }{c^{1-\nu}\ell} \,,\quad \nu = 2a-7=0.38\,,\quad \ell = 10 {\rm\,m}\,,
\end{aligned}
\end{equation}
having used $V_0^2=N/(32\rho_0)$, $c\simeq3$ and $A= E b^2\rho_0f m_\ast N_0 /
(\pi^3 N k_\ast^{(1-\nu)/2}) $ -- prefactor of the modified Garrett and
Munk spectrum \eqref{eq:ModifiedGMspectrum}, taking into account the
change of variables to isopycnal coordinates and to $k-m$
space. The following values of the physical parameters are used: $E_0=6.3\times
10^{-5}$ is the \gd{GM76} non-dimensional {\it energy level}, $b=1300$ m,
$\ell=10$ m, $\rho_0=1000$ kg$/$m$^3$, $N_0=0.00524$ s$^{-1}$, $f =
2\cdot 7.3\times 10^{-5} \sin(l)$ (at latitude $
l=32.5^\circ$). Moreover, we have used $m_\ast=4\pi N/(b N_0)$ and
$k_\ast=c m_\ast f/N$, with $c=3$. The factor $c$ has been added as a normalization factor to ensure that the energy level of the modified GM spectrum preserves the same energy level of the original GM76 spectrum, in an effort to minimize arbitrariness in the choice of $k_\ast$. Now, we can make an estimate of the dissipated
power at high wave numbers, \gd{using $N=N_0$ and $f=f_0$}, which gives:
\begin{equation}\label{eq:pownonlocnum}
\begin{aligned}
&\mP_{\rm out, h} \simeq -3.8 \times 10^{-9}\, \text{\rm W  kg}^{-1}\,,\\
&\mP_{\rm out, v} \simeq - 5.2 \times 10^{-9}\, \text{\rm W  kg}^{-1}\,.
\end{aligned}
\end{equation}
This amounts to a total dissipated power
\begin{equation}\label{eq:Ptot}
	\mP_{\rm out} = \mP_{\rm h} + \mP_{\rm v} \simeq -9.0 \times 10^{-9}\, \text{\rm W kg}^{-1}\,.
\end{equation}
The integration along the boundaries leading to $\mP_{\out,\h}$ and
$\mP_{\out,\v}$ is represented in the schematic in Fig.~\ref{fig:0}
and will be discussed in Sec.~\ref{sec:TheEnd}. Furthermore, we
are able to decompose both horizontal and vertical transfers
into local and scale separated sub-contributions, each of which is
dominated by a particular type of interactions: the ID process
dominates the scale-separated interactions while the local
interactions are dominated by triads that are quasi-colinear in the
horizontal plane. This will be illustrated in Sec.~\ref{sec:WKE}.

Surprisingly, the ID concept on which much of the understanding of
internal wave interactions is based, turns out to be quite marginal in
the economy of the total energy fluxes. However, as will be shown in
Sec.~\ref{sec:ID}, its analytical tractability turns out very useful
for the interpretation of the direction of the energy cascade through
scales.

Formulas~\eqref{eq:Ptot} and~\eqref{eq:powtot} \gd{can be compared
directly with the result of the finescale parameterization with the
same input spectrum},
Eq.~\eqref{eq:ModifiedGMspectrum}, the GM76 spectrum modified in such
a way that the high wavenumber power-law behavior matches that of the
stationary solution of the wave kinetic equation, Eq.~\eqref{eq:powlaw}. As
outlined in the introduction, use of the finescale parameterization
formula~\eqref{eq:FinescaleParameterization}
\citep{polzin1995finescale} for such spectrum yields
\begin{equation}
    \mP_{\rm finescale} \simeq 5.9\times 10^{-9}
    \,\text{W}/\text{kg}\,.
\end{equation}
The consistency between the finescale parameterization and the first-principles results~\eqref{eq:powtot},~\eqref{eq:pownonlocnum} and~\eqref{eq:Ptot} will be discussed in Sec.~\ref{sec:TheEnd}.

\section{The Internal-Wave Kinetic Equation and its steady state}\label{sec:WKE}

The hypotheses that have been made are the following.  We consider a
vertically stratified, spatially homogeneous oceanic internal
wavefield, expressed in isopycnal coordinates in the non-rotating,
hydrostatic approximation. The \gd{vertical stratification gradient} profile is assumed to be
constant and the wavefield isotropic in the horizontal directions. We
also assume vertical isotropy, i.e. symmetry $m\leftrightarrow -m$, so
that the description can be restricted to the vertical wavenumber
magnitude $m$ requiring that
{{$n(\bk,m)=n(\bk,+|\bm|)+n(\bk,-|\bm|)=2n(\bk,\bm)$}} (standard
one-sided {\it vs} two-sided spectrum definition on a symmetric
domain). The non-rotating, hydrostatic dispersion relation is given by
Eq.~\eqref{eq:disp}. Finally, we assume zero potential vorticity.

Let us consider an ensemble of random internal waves, in the joint
limit of large box and weak nonlinearity \gd{(\cite{ZLF,X3,NazBook,eyink2012kinetic,chibbaro20184,deng2021full})}.  Under
the above assumptions, the following wave kinetic equation, describing the
time evolution of the 3D wave-action spectrum, is derived
(\cite{LT,LT2,iwthLPTN}):
\begin{equation}\label{eq:kineq}
\begin{aligned}
	&\frac{\partial n(\bp)}{\partial t} = \mI(\bp)\,,\qquad \text{with}\\
	&\mI(\bp)= \frac{8\pi}{k}\int \Big(
        f^0_{12} |V^0_{12}|^2\delta_{\bm-\bm_1-\bm_2}
        \delta_{\omega_0-\omega_1-\omega_2} \frac{k k_1
          k_2}{\Delta_{012}}\\
          &\qquad\qquad\quad- (0\leftrightarrow1) -
        (0\leftrightarrow2)\Big) d k_1 d k_2 d\bm_1d\bm_2\,,
\end{aligned}
\end{equation}
%
where the term $f^0_{12}=n(\bp_1)n(\bp_2)-n(\bp)(n(\bp_1)+n(\bp_2))$
contains the dependence on the spectrum, double-ended arrows indicate the permutation of indices, and $V^0_{12}$ is the matrix
element quantifying the magnitude of the nonlinear interactions
between the triad of wavenumbers $\bp$, $\bp_1$ and $\bp_2$.
\begin{figure}
\begin{centering}
\includegraphics[width=\linewidth]{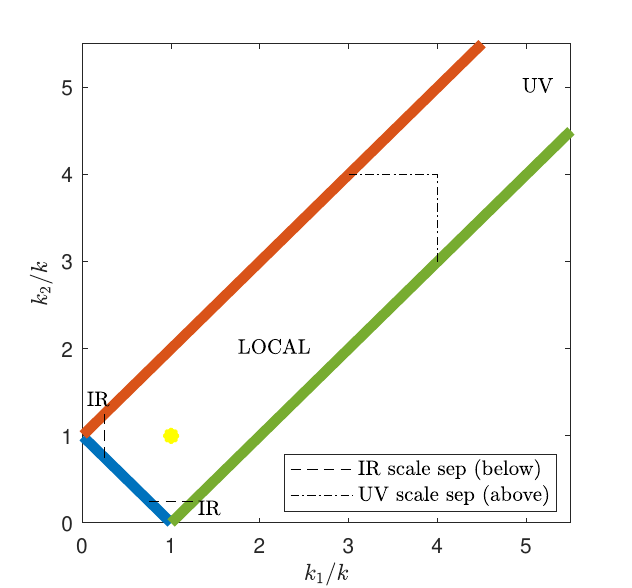}
\includegraphics[width=\linewidth]{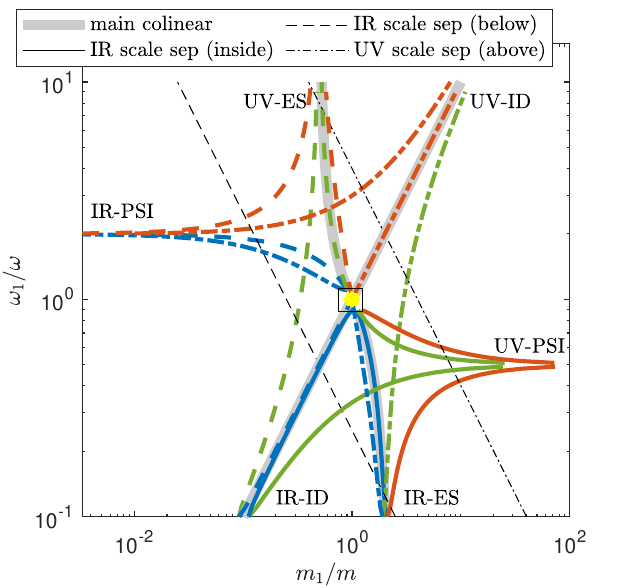}
  \caption{{\bf Top}: The
    resonant manifold in the space $k_1-k_2$, also referred to as the
    {\it kinematic box}, delimited by the colored {\it colinear}
    boundaries, where the three horizontal wavenumbers of each
    interaction are colinear. The corners of the box at points $(0,1)$
    and $(1,0)$ are the regions with (`infrared') extreme scale
    separation, where the ID scattering gives the leading-order
    contribution. {\bf Bottom}: The resonant manifold in the space
    $\omega_1-m_1$. Each of the colinear boundaries in the upper panel
    maps into six distinct curved edges of the same color,
    respectively. The result is a resonant manifold made of six
    lobes. Two of the lobes contribute to the ID leading order
    contribution in the scale separated region (labeled by $\rm
    IR-ID$), involving the interaction of a small wavenumber $\bp_1$
    that induces the scattering between the two much larger
    wavenumbers $\bp$ (yellow dot in the plot) and $\bp_2$ (inside the
    square surrounding the yellow dot).
  The separation between the scale separated and the local regions in the two plots are intended for a delimiting value of $\epsilon = 1/16$.}
 \label{fig:2}
\end{centering}
\end{figure}
\begin{figure}
\begin{centering}
\includegraphics[width=\linewidth]{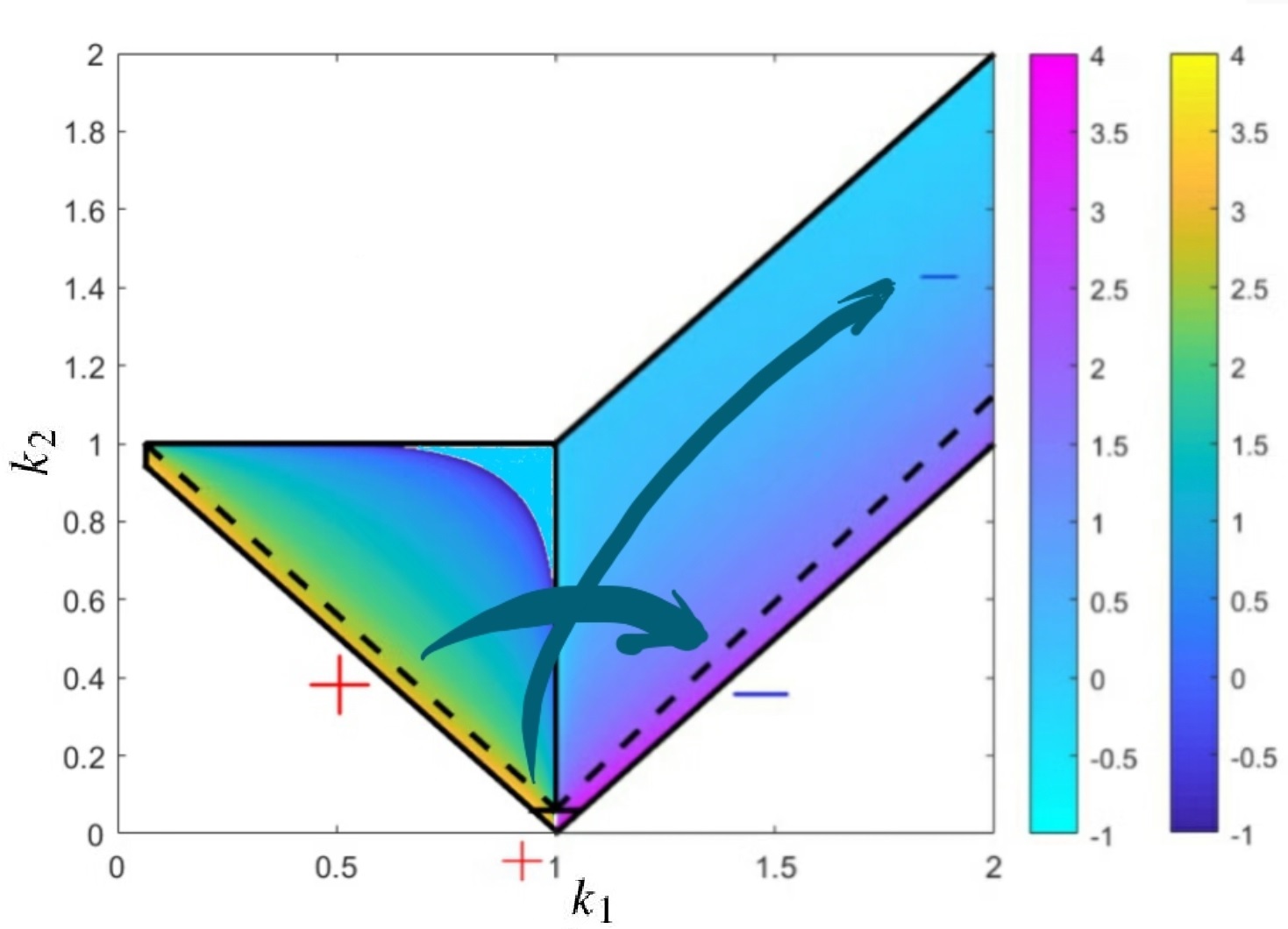}
\includegraphics[width=\linewidth]{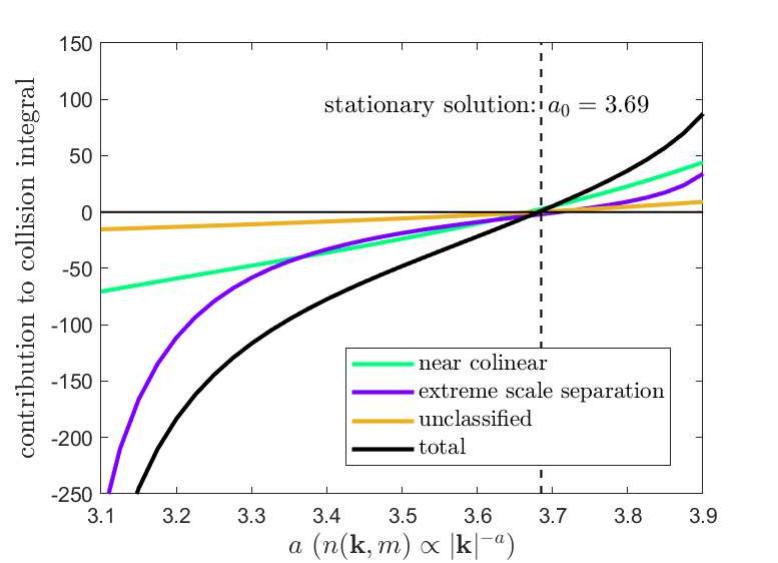}
  \caption{Both figures are from
    \cite{dematteis_lvov_2021}. {\bf Top}: representation of the
    magnitude of the interactions (integrand of Eq. \eqref{eq:kineq})
    for triads with horizontal wave numbers $k=1$, $k_1$ and $k_2$,
    for the stationary solution $(a,b)=(3.69,0)$. The small triangle
    above $k_1=1$, $k_2=0$ is the ID-dominated region. The thin
    regions delimited from above by a dashed line are the
    near-colinear regions. The colormaps represent the base-10
    logarithm of the magnitude of the contribution, while the
    left/right colormap denotes negative/positive contributions,
    respectively. The arrows depict the stationary balance between
    different regions (whose overall sum has to be zero for
    stationarity), highlighting the downscale direction of the
    horizontal flux. {\bf Bottom}: On the segment $a\in(3,4)$,
    $b=0$, breakdown of the contributions to the collision integral as
    sum of the scale-separated region, dominated by ID, and the more
    local colinear region. The rest of the unclassified triads gives a
    subleading contribution. The total vanishes for $a\simeq3.69$, the
    stationary state of the internal wave kinetic equation~\eqref{eq:kineq}.}
 \label{fig:3}
\end{centering}
\end{figure}
The
calculation of this interaction matrix element is challenging, with
early expressions given by
\cite{Olbers1973,Voronovich,milder1982hamiltonian,caillol2000kinetic}. In
the current manuscript we use the interaction matrix
element computed in \cite{LT2} by using the Hamiltonian formulation of
\cite{LT}. In Eq.~\eqref{eq:kineq}, the two delta functions impose the
conservation of vertical momentum and energy in each three-wave
interaction. The factor $\Delta_{012}$ comes from analytical
integration of the horizontal momentum delta function and is
proportional to the area of the triangle with sides $k$, $k_1$ and
$k_2$.  The nonlinear collision integral $\mI(\bp)$ contains all of
the information about the nonlinear resonant energy transfers
involving point $\bp$ in Fourier space, after integrating out the
azimuthal angle thanks to horizontal isotropy.

The two independent delta functions can be integrated over reducing
the domain of the integrand to the {\it resonant manifold}, with two
degrees of freedom left. In Fig.~\ref{fig:2} two equivalent
representations of the resonant manifold are shown, in the $k_1-k_2$
space (upper panel) and in the $m_1-\omega_1$ space (lower panel). In
the $k_1-k_2$ space, the triangular inequalities constrain the
possible interactions to the so-called {\it kinematic box}, delimited
by the colored boundaries in the figure. The points on these three
boundaries identify triads with colinear horizontal wavenumbers. The
infrared (IR) scale-separated interactions, where ID dominates, are
delimited by a dashed line at $k_1=\epsilon$ or $k_2=\epsilon$ ({\it
  cf.} Fig.~\ref{fig:1}, upper panels). An equivalent representation
of the resonant manifold has $m_1$ and $\omega_1$ as the two
independent degrees of freedom, as represented in the bottom panel of
Fig.~\ref{fig:2}. The result is a resonant manifold made of six lobes.
Each of the colinear boundaries in the upper panel maps into six
distinct curved edges of the same color, respectively\gd{, in the lower panel}.  In the
$m_1-\omega_1$ space, the IR scale-separated region is mapped into the
part of the resonant lobes to the left of the dashed line (if $\bp_1$
is the small wavenumber or small frequency in the interaction and
$\bp_2\simeq \bp$) or into the small box surrounding the yellow dot
(if $\bp_1\simeq \bp$, where the yellow dot denotes $\bp$, and
$\bp_2\ll\bp$). By the ID asymptotics, in the $m_1$ coordinate the
width of such box is constrained to be roughly the interval
$[1-\sqrt{\epsilon},1+\sqrt{\epsilon}]$ (\gd{see Appendix B and} {\it cf.} Fig.~\ref{fig:1},
lower panels).


In the non-rotating limit of Eq. \eqref{eq:kineq} all of the factors
in the integrand of $\mI(\bp)$ are power laws in the variables $k$ and
$m$ and therefore it is natural (and general) to restrict the possible
stationary solutions to a power law of the form
\begin{equation}\label{eq:powlawb}
    n(\bp)=Ak^{-a}m^{-b}\,.
\end{equation}
This allows us to represent the possible solutions in the 2D power-law
plane $a-b$ and obtain analytical results that could not be pursued
otherwise.  Using the scale invariant properties of $\mI(\bp)$ and the
ansatz \eqref{eq:powlawb}, Eq. \eqref{eq:kineq} in stationary
conditions reduces to
\begin{equation}\label{eq:IDeqpowlaw}
\begin{aligned}
 \frac{\partial n(k,m)}{\partial t} = \frac{4\pi}{\gamma} (A V_0)^2 k^{-2a+5} |m|^{-2b+1} I(a,b)=0\,,
 \end{aligned}
\end{equation}
expressed for the 2D action spectrum.  Here, $V_0$ is a dimensional
constant prefactor of the matrix elements and
$I(a,b)=\mI(k=1,m=1)\gamma/(A V_0)^2$ is the non dimensional collision
integral: it is a function of $a$ and $b$ only, that must vanish in
order for the solution to be stationary. It has been shown
\citep{iwthLPTN,dematteis_lvov_2021} that $I(a,b)$ is a finite
(non-divergent) integral only on the segment $a\in(3,4),
b=0$. Moreover, on such convergence segment one finds that $I(a,b)$
vanishes at $a\simeq3.69,b=0$, which represents the only well-defined
stationary solution to Eq. \eqref{eq:kineq}. This is shown in the
bottom panel of Fig.~\ref{fig:3}, where the separate contribution of
the scale separated and local regions is made apparent, for different
values of $a\in(3,4)$ and $b=0$. In particular, we notice that among
the local interactions, those with quasi-colinear horizontal
wavenumber give the largest contribution. In the upper panel of
Fig.~\ref{fig:3}, we show the magnitude of the integrand for the
stationary solution, in the kinematic box. The quasi-colinear regions
are delimited by dashed lines and the integrand is there visibly much
larger than in the rest of the box. Therefore, the local contribution
is mainly given by triads close to horizontal colinearity, meaning
that in three dimensions the three members of the triad $\bp$,
$\bp_1$, $\bp_2$ lie on the same vertical plane. As far as the local
interactions are concerned, the results presented in
Sec.~\ref{sec:loc-extr} are obtained by numerical recursive
integration in suitable regions of the kinematic box, whose result is
illustrated in Fig.~\ref{fig:1}\gd{, with the same numerical method used by \cite{dematteis_lvov_2021}}. The arrows in the upper panel of
Fig.~\ref{fig:3} symbolize the action fluxes between the waves of a
triad $\bp,\bp_1,\bp_2$: if the integrand at point $k_1,k_2$ is
positive, $\bp$ is `created' in the interaction, and this contributes
to an increase of its content of action in time; if the integrand is
negative, $\bp$ is `absorbed' in the interaction, and its action
content is depleted. Eq.~\eqref{eq:kineq} has intrinsic turbulent
character, and so does its stationary state: it is a nonequilibrium
solution with a flux of energy across scales that is constant in time
and directed downscale \gd{(toward larger values of $k$)}.

It is worth mentioning that the introduction of a minimal frequency
equal to the inertial frequency $f$, a maximal frequency equal to the
buoyancy frequency $N$, and of physical cutoffs at small and large
vertical spatial scales has a chance to regularize the collision
integral also for spectra outside the convergence segment. A detailed
and comprehensive analysis of this issue is subject of current
research. 

\section{Induced Diffusion revisited}\label{sec:ID}

Although general, the integral formulation~\eqref{eq:int_cons} may be
hard to visualize. Further simplification of the picture may be
achieved by assuming that the transfer is dominated by triads with
extreme scale separation. In other words, in the decomposition of
Eq.~\eqref{eq:kineq_decomp} one assumes that $\mI^{({\rm
    sep})}\gg\mI^{({\rm loc})}$, so that $\mI\simeq \mI^{({\rm
    sep})}$, restricting the integration of the r.h.s. of
Eq.~\eqref{eq:kineq} to the IR corners of the kinematic box.  Since
the early works \cite{McComas1977,mccomas1981time,Muller86}, these
scale-separated interactions have been classified under the three
processes of Parametric Subharmonic Instability (PSI), Elastic
Scattering (ES) and Induced Diffusion (ID). In particular,
\cite{mccomas1981dynamic} interpreted the GM76 spectrum as resulting
from the stationary balance of a Fokker-Planck equation for the wave
action, derived under the assumption that the ID process dominates the
transfers. The ID process involves a net exchange of energy between
two almost identical wavenumbers mediated by a much smaller
wavenumber. In the simplest formulation of the ID theory, the
attention is focused on the large wavenumbers and the
small-wavenumber, low-frequency part of the spectrum (the so-called
near-inertial region) is considered as a decoupled independent
reservoir that is given and constant in time. In the system of large
wavenumbers alone, then, one notices that ID implies the scattering
between two neighboring wavenumbers. Neglecting the fact that the
scattering would not occur without the mediation of the smaller
wavenumber reservoir, this process preserves wave action in the
high-wavenumber region.  Note that the wave action can also be
interpreted as the `number of quasi-particles' (or waves), and here
one wave is scattered into another one locally preserving the total
`number of waves'. The ID equation derived in \cite{McComas1977} is given by Eq. \eqref{eq:IDrhs}-\eqref{eq:kineq_decomp}, setting $\mI^{({\rm loc})}=0$,
where the $a_{ji}$'s ($i,j=1,2,3$) denote the coefficients of the
diffusion tensor, with {{explicit}} expressions provided in the
    appendix therein. For a simple visualization of the energy flux, here we use a representation in the 2D plane
    $k-m$ (or equivalently $\omega-m$).  Using the \gd{transformation} in
    cylindrical coordinates, horizontal isotropy and vertical
    isotropy, the ID equation for the 2D action density
    gives
\begin{equation}\label{eq:IDeq}
\begin{aligned}
    &\frac{\partial n(k,m)}{\partial t} = -\nabla \cdot \bJ^{(n)}(k,m)\,,\\
    &\bJ^{(n)}(k,m)=\left( \left(\frac{a_{kk}}{k} - a_{kk}\frac{\partial}{\partial k} - a_{km} \frac{\partial}{\partial m}\right)\right.\,,\\ &\,\quad\qquad\qquad\left.\left(\frac{a_{mk}}{k} - a_{mk}\frac{\partial}{\partial k} - a_{mm} \frac{\partial}{\partial m} \right)\right)n(k,m),
    \end{aligned}
\end{equation}
where $\nabla = (\partial/\partial k,\partial/\partial m)$ and
$a_{kk}=a_{11}=a_{22}$, $a_{km}=a_{mk}=a_{13}=a_{23}=a_{31}=a_{32}$,
$a_{mm}=a_{33}$. The effects of $a_{12}$ and $a_{21}$ are here
cancelled by assuming horizontal isotropy. Notice that the 3D action diffusion coefficients contribute to both
advection and diffusion terms for the 2D action in
Eq.~\eqref{eq:IDeq}.

We would like to stress two further points. First, Eq.~\eqref{eq:IDeq}
is for the wave action density and not for the energy density because
in the high-wavenumber part of the spectrum \gd{it is action, not energy, to be} conserved in
the ID picture, as explained above. By making the change of variables
$e(k,m) = \omega n(k,m)$ one concludes that expressing the same
equation for the energy density implies the presence of an extra
energy source/sink term that accounts for the absorption/creation of
the member of the triad in the near-inertial reservoir, whose energy
is transferred nonlocally to/from the high wave number region -- a
graphical representation of this fact is found e.g. in Fig. (6) of
\cite{mccomas1981dynamic}. For this reason Eq.~\eqref{eq:IDeq} is
preferably expressed for the action, but one can obtain the energy
flux simply by using $\bJ^{(e)}(k,m)=\omega \bJ^{(n)}(k,m)\,$.

Second, we stress that equations~\eqref{eq:kineq} and~\eqref{eq:IDeq}
are not equivalent, as made clear in Sec.~\ref{sec:loc-extr}. The latter is derived from the former under the
assumption that all of the energy transfers are scale separated and
neglecting the rest of the interactions. This is going to be analyzed below.

\subsection{Closure for the ID energy flux: non-rotating case}\label{sec:IDa}

Now, for the Fokker-Planck equation~\eqref{eq:IDeq} to have the correct scale-invariant properties of
Eq.~\eqref{eq:IDeqpowlaw}, at the stationary state, the following
consistency conditions must hold for the coefficients of the diffusion
tensor:
\begin{equation}\label{eq:diff}
\begin{aligned}
    &a_{kk}=c_{kk} k^{6-a}m^{1-b}\,,\quad a_{km}=c_{km} k^{5-a}m^{2-b}\,,\\ &\qquad\qquad a_{mm}=c_{mm} k^{4-a}m^{3-b}\,,
    \end{aligned}
\end{equation}
where the $c_{ij}$'s are constants that in principle can be determined
by straightforward calculation. For instance, explicit expressions of
$c_{kk}$ and $c_{mk}$ for the steady state are given in
Eq.~\eqref{eq:coeff_diff}.  The scalings in Eq.~\eqref{eq:diff} are a
consequence of the non-rotating assumption, while Eq.~\eqref{eq:IDeq}
has the same form also in the presence of background rotation (\gd{the rotating case will be considered in Sec.~4\ref{sec:IDb}}). In Eq.~\eqref{eq:kineq},
the convergence conditions that technically restrict the range of
possible solutions onto the convergence segment $a\in(3,4),b=0$ are
due to the singularity in the ID limit. Thus, the same considerations
should be applied for the well-posedness of the
coefficients~\eqref{eq:diff}. Since \cite{McComas1977,
  mccomas1981time} the Fokker-Planck equation has been shown to enjoy
stationary states for all points on the two lines $b=0$ and
$b=3-2a/3$. This has been re-derived in \cite{iwthLPTN}, highlighting
how the result is based on a {\it restriction to the limit of the
  infrared ID interactions}. Despite this, we find that for $b=0$
there are exact cancellations between the ID leading order of the
singularities of the collision integrand (r.h.s. of
Eq.~\eqref{eq:kineq}), that need to be treated with particular
care. An exact balance between the leading non-zero ID contributions,
{\it both infrared and ultraviolet}, allowed
\cite{dematteis_lvov_2021} to obtain analytically that the ID solution
is stationary, independent of the other interactions, for $a\simeq
3.69$, which is compatible with the full balance obtained with all
interactions. This can be observed in the lower panel of
Fig. \ref{fig:3}, \gd{where the balances between scale-separated interactions and local interactions are shown to vanish at $a\simeq
3.69$ separately}. Therefore, at least for $b=0$, the Fokker-Planck
equation \eqref{eq:IDeq} enjoys the same stationary state $(3.69,0)$
as Eq. \eqref{eq:kineq}, while the other states with $b=0, a\neq3.69$
are found to be (at a subleading order \gd{that had been neglected in previous works}) off-balance.

Using the expressions \eqref{eq:diff} in Eq. \eqref{eq:IDeq}, for a generic power-law spectrum \eqref{eq:powlawb} equivalent to a 2D action spectrum $n(k,m)=4\pi A k^{-a+1}m^{-b}$,  yields:
\begin{equation}
\begin{aligned}
    \frac{\partial n(k,m)}{\partial t} &= -4\pi A\frac{\partial}{\partial k}\left[\left( a c_{kk} + b c_{km}\right)k^{6-2a}m^{1-2b} \right]\\
    & \;\;\;-4\pi A \frac{\partial}{\partial m}\left[\left( a c_{km} + b c_{mm}\right)k^{5-2a}m^{2-2b} \right]\\
    &= 4\pi A\left[(2a-6)\left( a c_{kk} + b c_{km}\right)\right.\\
    &\qquad\;\;\; \left.+ (2b-2) \left( a c_{km} + b c_{mm}\right)\right]k^{5-2a}m^{1-2b}\,.
\end{aligned}
\end{equation}
Now, assuming that the given spectrum is stationary, the r.h.s. must vanish for all $k$ and $m$. This implies the condition:
\begin{equation}\label{eq:stat}
    \frac{a c_{kk} + b c_{km}}{a c_{km} + b c_{mm}} = \frac{2-2b}{2a-6}\,.
\end{equation}
Using again \eqref{eq:IDeq} and \eqref{eq:diff}, we obtain the following formula for the stationary energy flux,
\begin{equation}\label{eq:fluxen}
\begin{aligned}
	\bJ^{(e)}(k,m) & = 4\pi \gamma A \left( (a c_{kk}+b c_{km}) k^{7-2a} m^{-2b},\right.\\
	&\qquad\qquad\left.(a c_{km}+b c_{mm}) k^{6-2a} m^{1-2b}\right)\,,\\
&=C_0 \left((2-2b)k^{7-2a} m^{-2b},\right.\\
&\qquad\;\; \left.(2a-6) k^{6-2a} m^{1-2b} \right)\,,\quad C_0>0\,,
\end{aligned}
\end{equation}
where the last line is true if the solution is a stationary state
($I(a,b)=0$), ensuring the validity of the condition
\eqref{eq:stat}. Moreover, except for the overall normalization
constant $C_0$, this relation provides pointwise knowledge of the
steady state flux. This is used next to investigate the direction of the
steady state energy flux. As a consistency check on the results of Sec.~\ref{sec:loc-extr}, notice that for the steady state coefficients in Eq.~\eqref{eq:coeff_diff} we have: $5.8/4.0\simeq 2/(2a-6)\simeq 1.45$\,, verifying the condition~\eqref{eq:stat}.

We then consider the inertial
range as the region such that $m_{\min}<m<m_{\max}$ and
$f<\omega<N$ (see Fig. \ref{fig:0}), which due to the dispersion relation \eqref{eq:disp}
corresponds to a trapezoid in $k-m$ space. The dispersion relation
also allows us to change variables and express the flux in $\omega-m$
space, in which the inertial range is simply the
rectangle $[f,N]\times[m_{\min},m_{\max}]$. In these coordinates, the
energy flux \eqref{eq:fluxen} takes the form
\begin{equation}\label{eq:fluxen2}
\begin{aligned}
{\bJ^{(e)}}(\omega,m) &= C_0 \gamma^{2a-7} \left( (8-2a-2b)\omega^{7-2a} m^{7-2a-2b},\right.\\
&\left.\quad\qquad\qquad (2a-6) \omega^{6-2a} m^{8-2a-2b}\right)\,.
 \end{aligned}
\end{equation}
This result allows for transparent graphical interpretation of the
  nature and paths of the Fourier-space diffusion-like energy flows. Approximating the kinetic equation with the differential conservation
form \eqref{eq:IDeq} allows us to analyze the direction of the fluxes
within the ID paradigm. Eq. \eqref{eq:IDeq} is nothing but a
projection of the Fokker-Planck equation \eqref{eq:IDrhs} on the 2D
$k-m$ space.

\begin{figure}
\begin{centering}
\includegraphics[width=\linewidth]{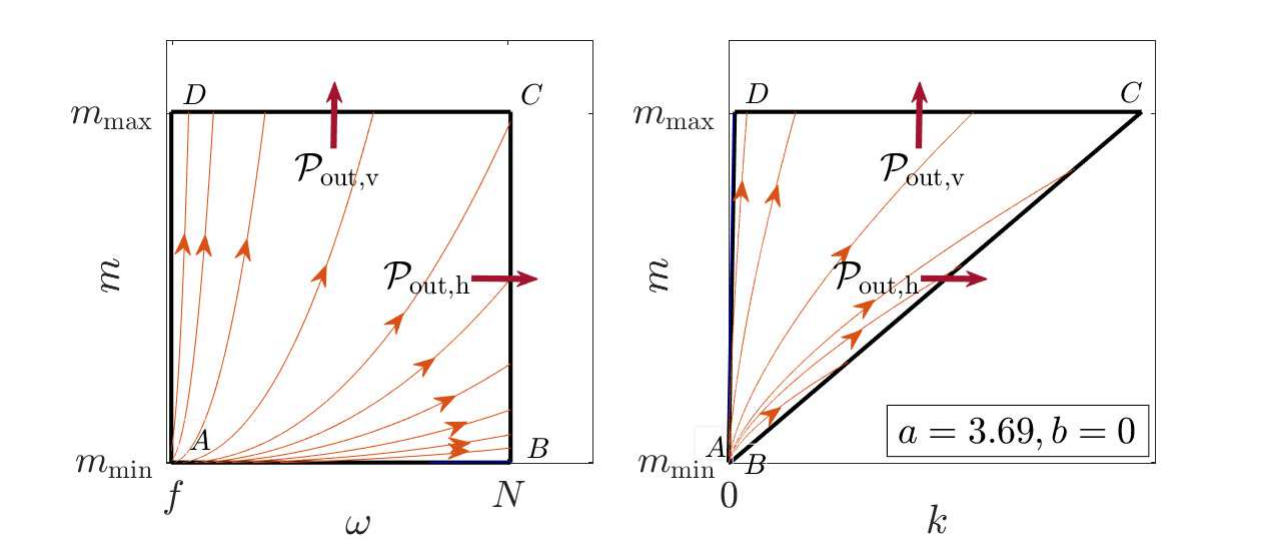}
\includegraphics[width=\linewidth]{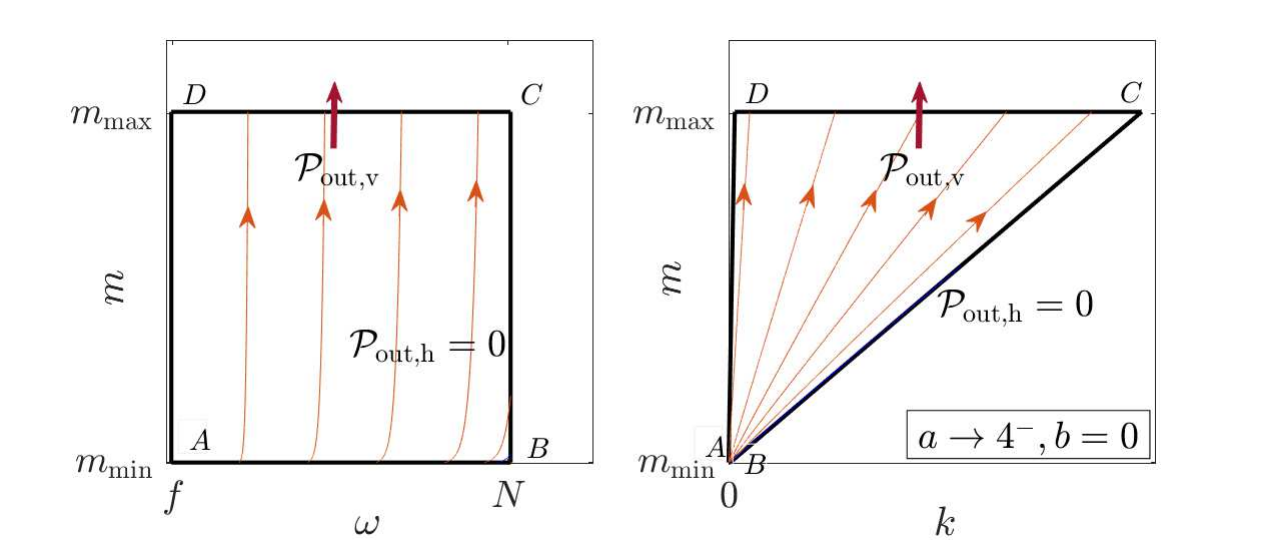}
\includegraphics[width=\linewidth]{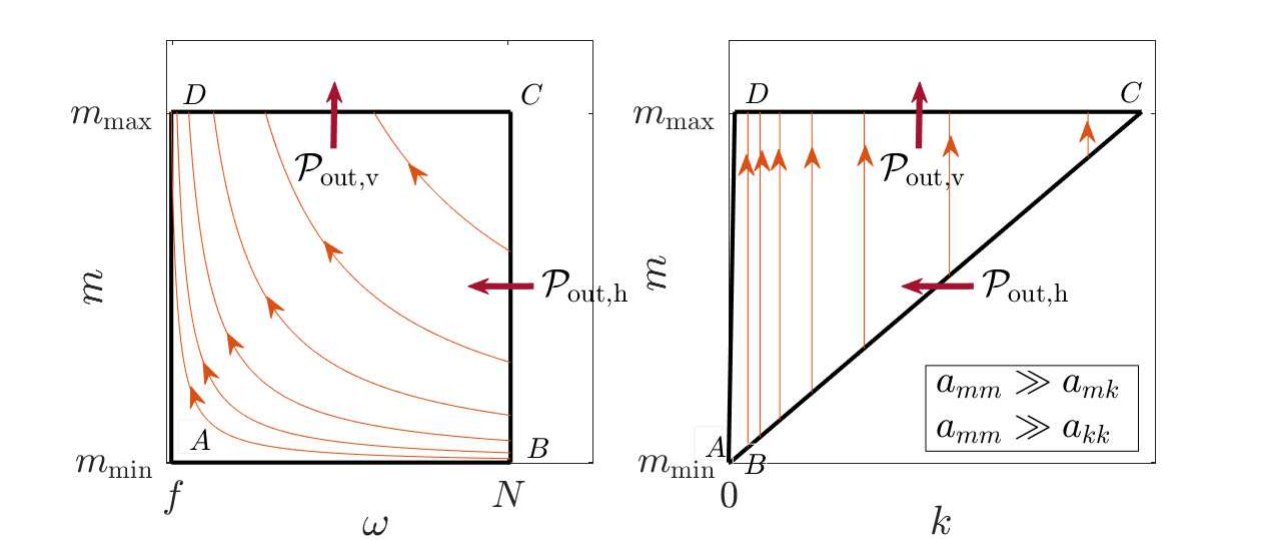}
  \caption{Direction of the energy flux as
    a function of the power-law exponent according to
    Eqs. \eqref{eq:fluxen}-\eqref{eq:fluxen2}. Top panels: $a=3.69,
    b=0$; central panels $a\to4^-, b=0$ (GM76 solution); bottom panels:
    constrained flux direction according to \cite{McComas1977}, after
    the vertical-vertical-diffusion-only assumption is made. Left
    panels are in $\omega-m$ space, right panels are in $k-m$ space; the two representations are equivalent.}
 \label{fig:4}
\end{centering}
\end{figure}

Now, a further simplification, proposed in \cite{McComas1977}, can be
made by asserting that the transfer is dominated by the
$a_{33}=a_{mm}$ term of the diffusion tensor $a_{ji}$. Below, we focus
on analyzing what this approximation entails, and we find that an
inverse cascade of energy in frequency is necessarily implied,
requiring existence of an energy source at high frequencies in order
to be sustained. On the other hand, for the stationary solution of the
wave kinetic equation we show that, if all components of the diffusion
tensor are considered, the Fokker-Planck equation leads to a cascade
of energy from low to high frequency.  These results are presented in
Fig. \ref{fig:4}. Namely, in the upper panel of Fig. \ref{fig:4}, we
show the streamlines of the energy flux in both systems of coordinates
-- Eqs. \eqref{eq:fluxen2} and \eqref{eq:fluxen}, respectively -- for
the stationary solution $a=3.69,b=0$. In the $\omega-m$
representation, the flux is downscale in both frequency and
vertical-wavenumber directions. Importantly, we observe that a source
of energy at low frequency and small vertical wavenumber would be
compatible with this flux. Considering the relative proximity of the
high-wavenumber GM spectrum in the space of power-law solutions, and
arguing that the effects of physical cut-offs may modify the
stationary solution toward the GM slope itself, we can observe how the
energy-flux streamlines behave as $a\to4$. We observe that the
streamlines change continuously in the parameters $a$ and $b$, tilting
toward the vertical direction in $\omega-m$ space, as $a\to4$. This is
depicted in the central panels of Fig. \ref{fig:4}. Although not
rigorous, this observation is in agreement with the downscale energy
cascade in the finescale parameterization paradigm
\citep{polzin2014finescale}, interpreted as an essentially vertical
process in $\omega-m$ space.

Since the coordinate systems considered have different units in the
vertical and horizontal direction, it is useful to quantify the flux
direction using integrated quantities that can be compared
directly. We thus compute the power flowing out of the fixed boundary
$BCD$, $\mP^{(e)}_{BCD}=\mP^{(e)}_{BC}+\mP^{(e)}_{CD}$, where the two
contributions are given by integration of the component of the flux
normal to the sides $BC$ and $CD$, respectively. The computation is
easiest in $\omega-m$ space, yielding:
\begin{equation}\label{eq:P}
\begin{aligned}
	{\mP^{(e)}_{BC}} &= \int_{m_{\rm min}}^{m_{\rm max}} d m \, {\bJ^{(e)}}(N,m)\cdot (-1,0) \\
	& = -C_0 \left(\frac{\gamma}{N}\right)^{2a-7} \left(m_{\rm max}^{8-2a-2b} - m_{\rm min}^{8-2a-2b}\right)\,,\\
	{\mP^{(e)}_{CD}} &= \int_{f}^{N} d\omega \, {\bJ^{(e)}}(\omega,m_{\rm max})\cdot (0,-1)\\
	&= -C_0 \frac{2a-6}{2a-7}\gamma^{2a-7}  m_{\rm max}^{8-2a-2b} \left( f^{7-2a} - N^{7-2a} \right)\,,
\end{aligned}
\end{equation}
with the convention that an outgoing/incoming power is negative/positive since it is lost/gained by the set under consideration (the box $ABCD$).
So, we define the ratio
\begin{equation}\label{eq:Rs}
\gd{R^{\rm (sep)}}=\frac{\mP^{(e)}_{CD}}{\mP^{(e)}_{BC}}= \frac{2a-6}{2a-7}\frac{\left(N/f\right)^{2a-7}-1}{ 1-\left(m_{\max}/m_{\min}\right)^{2(a+b-4)}}
\end{equation}
to characterize the global vertical-to-horizontal downscale energy
transfer ratio\gd{, restricted to the scale-separated interactions under scrutiny in the current section}. Substituting $a=3.69$ and $b=0$, we obtain $\gd{R^{\rm (sep)}}\simeq
4.5$: in the ID paradigm, the downscale flux in the vertical
direction is about a half order of magnitude larger than in the
horizontal direction. With the same caveats about regularization by
suitable cut-offs, stationarity and departure from scale invariance,
we observe that the GM limit would imply $\gd{R^{\rm (sep)}}\to\infty$, in agreement
with the verticality of the flux in such limit.
Indeed, furthermore for the action one finds that
$\mP^{(n)}_{BCD}+\mP^{(n)}_{BAD}=0$, independent of the values of $a$
and $b$, since action is conserved by Eq. \eqref{eq:IDeq}. For the
energy, on the other hand, we have:
\begin{equation}\label{eq:source}
\begin{aligned}
    &\mP^{(e)}_{BCD}+\mP^{(e)}_{BAD} \\
    &= \frac{C_0 \gamma^{2a-7}}{7-2a}
    \left(m_{\rm max}^{-2(a+b-4)} - m_{\rm
      min}^{-2(a+b-4)}\right)\left( f^{7-2a} -N^{7-2a}\right)\,,
      \end{aligned}
\end{equation}
which is negative for $a+b-4<0$\,. This coincides with fluxes toward
higher frequencies, for which horizontal transport in $\omega-m$ space
is downscale and $|\mP^{(e)}_{BCD}|>|\mP^{(e)}_{BAD}|$. An action-conserving flux
toward larger frequencies necessarily
implies an energy increase. This does not violate energy conservation
nor the stationary balance! Simply, in the ID picture the extra energy
that appears at high wave numbers comes from the near-inertial
reservoir that acts as a nonlocal energy source in the continuity
equation for the energy density. This fact was explained in Figs. 5
and 6 of \cite{mccomas1981dynamic}, where they had in mind a flux toward smaller
frequencies implying a sink rather than a source at high
    frequency. For $a+b-4=0$ (which includes the GM76 case), instead,
$|\mP^{(e)}_{ABC}|=|\mP^{(e)}_{BCD}|$ since the flux is vertical in
$\omega-m$ space, i.e. action is transferred at constant $\omega$.

In \cite{McComas1977}, after deriving the Fokker-Planck equation, a further approximation is made by assuming that the
transfers are dominated by the $a_{33}$ element of the diffusion
matrix. This approximation is then discussed and analyzed further in
\cite{mccomas1981time,Muller86}. In the framework developed above, this
assumption is equivalent to setting $c_{mm}\neq0$ , and
$c_{kk}=c_{km}=0$.

Then, since the only non-zero
element is $c_{mm}$,
the energy flux in Eq. \eqref{eq:fluxen} is purely vertical in $k-m$ space {\it
  independent} of the values of $a$ and $b$. This is shown in the
bottom-right panel of Fig. \ref{fig:4}, representative of the ID picture
of \cite{mccomas1981dynamic}. As shown in the bottom-left panel, this
translates into an inverse cascade in frequency when transfers are
looked at in $\omega-m$ space. As pointed out in the introduction,
this fact has represented the first problem of the Oceanic
Ultraviolet Catastrophe, since a major energy source at high frequency
is believed not to be physically plausible.

Now, let us focus the attention on the case $b=0$. Looking at the
first line of Eq. \eqref{eq:fluxen}, for $b=0$ the approximation that
$c_{kk}$ and $c_{km}$ are negligible with respect to $c_{mm}$ appears
to be singular: since the factor $b=0$ makes the contribution of
$c_{mm}$ vanish, one has to look at the other terms that could give
finite contributions. In particular, according to
Eq. \eqref{eq:fluxen} (and keeping in mind the relations
\eqref{eq:stat}), in the $b=0$ case the horizontal flux is due to the
$a_{kk}$ diagonal element, whilst the vertical flux is due to the
$a_{mk}$ off-diagonal element of the diffusion matrix ({\it cf}
Eq. \eqref{eq:coeff_diff}). Notice that this consideration is only
based on the fact that $b=0$, and therefore it extends also to the GM
solution.

\begin{figure}
\begin{centering}
\includegraphics[width=\linewidth]{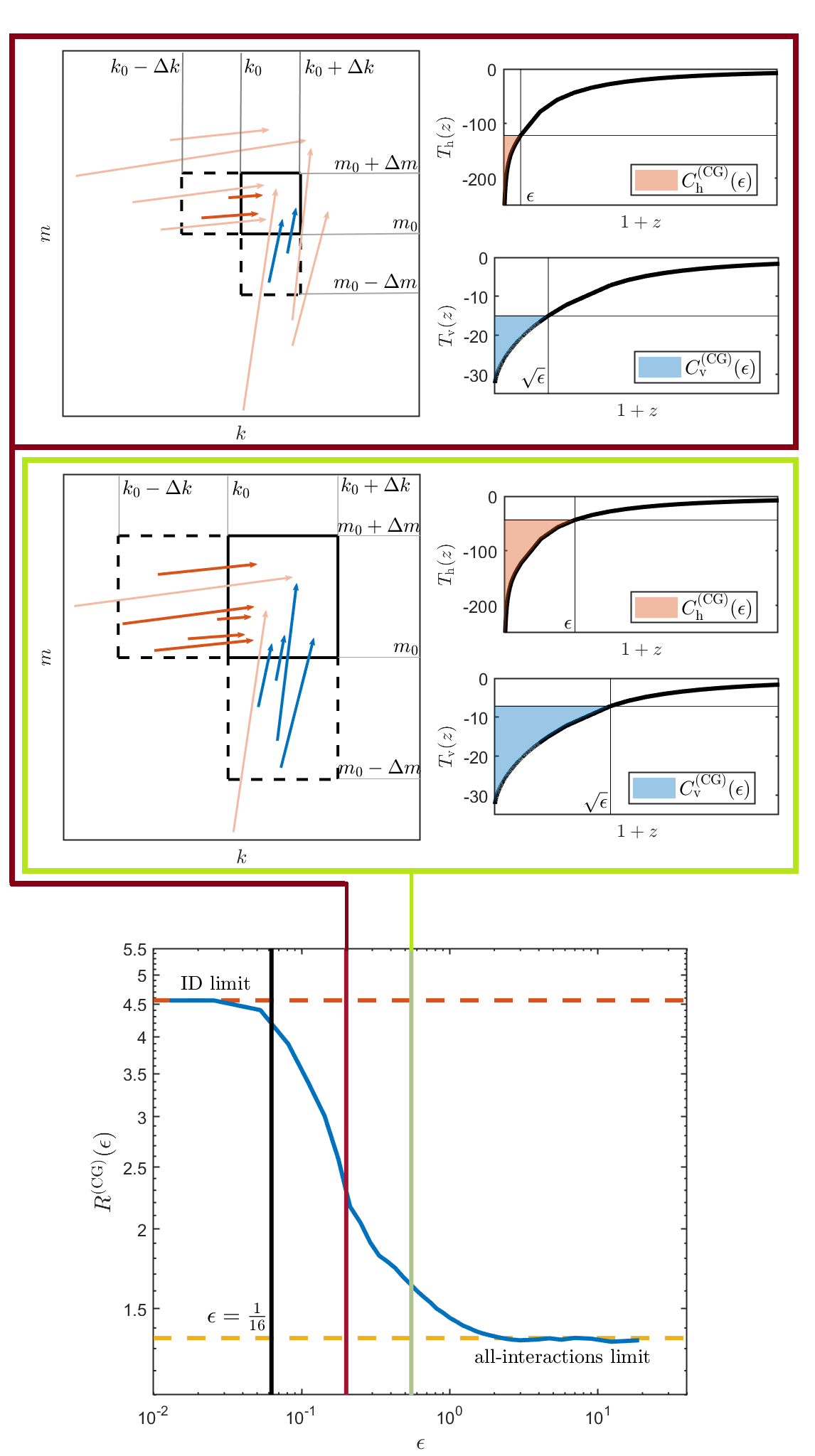}
  \caption{The value of the
    vertical-to-horizontal outgoing power ratio for a coarse-graining
    box with sides determined by $\epsilon$, defined as the ratio between
    Eq. \eqref{eq:CGpowB} and Eq. \eqref{eq:CGpowA}. For large $\epsilon$ the control box
    corresponds to the whole inertial range and all interactions
    are included\gd{, reproducing the ratio of the powers in Eqs.~\eqref{eq:powtot}-\eqref{eq:pownonlocnum} (dashed yellow line)}. As $\epsilon\to0$, more and more interactions are filtered out and the ratio becomes larger until it
    reaches quantitative agreement with the ID theory \gd{(Eqs.~\eqref{eq:P}-\eqref{eq:Rs}, dashed red line)} when the scale-separated
    interactions alone are left in the box.}
 \label{fig:5}
\end{centering}
\end{figure}

In Sec. \ref{sec:loc-extr}, these analytical results in the scale-separated region have successfully complemented the numerical results obtained for the local interactions. On the one hand, this has made it clear that assuming $I^{({\rm loc})}$ negligible with respect to $I^{({\rm sep})}$ is not justified. On the other hand, the fact that a non-negligible subset of interactions {\it are} diffusive provides direct knowledge of the pointwise diffusive part of the energy flux (see Figs. \ref{fig:4}-\ref{fig:5}) and allows us to draw important considerations for the pathways of energy.

For the stationary solution with
$a=3.69, b=0$, in particular, considering all terms of the diffusion matrix
has implied the {\it non-zero} flux \eqref{eq:fluxen2}, which is downscale both in
frequency and vertical wavenumber and is consistent with the steady state. Moreover, we have estimated vertical transport to
exceed horizontal transport by almost half order of magnitude in the
ID paradigm, meaning the off-diagonal element of the diffusion tensor
plays a leading role that had remained mostly undetected so far. The key to the solution of the  long-standing
paradoxes of the Oceanic Ultraviolet Catastrophe, according to to our
results, is thus to be found in non-negligible effects of previously neglected elements of the diffusion tensor.

The analytical results presented in this section can be made rigorous; this will be the subject of a companion paper. An intuitive picture goes as follows. Let us consider a squared partition of the inertial range in boxes of sides $\Delta k, \Delta m$, as represented for two different choices of $\Delta k, \Delta m$, in Fig. \ref{fig:5}.
Once a partition is fixed, let us define a coarse-grained model for which energy can be exchanged only through adjacent boxes in the partition, cutting off the rest of the interactions. We define the coarse-grained transfer integrals
\begin{equation}
\begin{aligned}
    &C_\h^{(CG)}(\epsilon) = \int_0^{\epsilon} dz \int_{z}^{\epsilon} dt\, \partial_zT_\h(t)\,,\\
    &C_\v^{(CG)}(\epsilon) = \int_1^{\sqrt{\epsilon}} dz\, \int_z^{\sqrt{\epsilon}} dt\,\partial_zT_\v(t)\,,
    \end{aligned}
\end{equation}
which tend to $C_\h$ and $C_\v$ for large $\epsilon$ and tend to
restrict the coarse-graining rectangular box to the ID region as
$\epsilon\to0$, with the correct scaling that relates the horizontal
side ($1-(1+\epsilon)^{-1}$) to the vertical side
($1-(1+\sqrt{\epsilon})^{-1}$) for the ID interactions. In agreement with
Eq. \eqref{eq:transfer}, we define coarse-grained powers exiting the inertial range, that relate
to the coarse-grained transfer integrals via
\begin{equation}\label{eq:CGpowA}
\begin{aligned}
    \mP_{\out,\rm h}^{(\rm CG)}(\epsilon) = &\,4\pi
  \frac{(NV_0A)^2}{(8-2a)g} \left(\frac{N}{\gamma}\right)^{7-2a}\\
  &\;\;\;\times(m_{\rm max} - m_{\rm
    min})^{8-2a}\, C_{\rm h}^{(\rm CG)}(\epsilon) \,,
    \end{aligned}
    \end{equation}
\begin{equation}\label{eq:CGpowB}
\begin{aligned}
    \mP_{\out,\rm v}^{(\rm CG)}(\epsilon) =&\,4\pi \frac{(NV_0A)^2}{(2a-7)g} 
  \left[\left(\frac{f}{\gamma}\right)^{7-2a}-\left(\frac{N}{\gamma}\right)^{7-2a}\right]\\
  &\;\;\;\times m_{\rm
      max}^{8-2a}\,C_{\rm v}^{(\rm CG)}(\epsilon) \,,
      \end{aligned}
\end{equation}
Let us define the ratio $R^{(\rm CG)}(\epsilon)=\mP_{\out,\rm v}^{(\rm
  CG)}(\epsilon)/\mP_{\out,\rm h}^{(\rm CG)}(\epsilon)$. For large values of $\epsilon$, the coarse-grained model includes all interactions and Eqs. \eqref{eq:CGpowA}-\eqref{eq:CGpowB} reduce to Eq. \eqref{eq:transfer}; therefore, using Eq. \eqref{eq:pownonlocnum}, we have: $R^{(\rm CG)}(\epsilon)\to 5.2/3.8\simeq 1.4$ for large $\epsilon$. On the other hand, as $\epsilon$
is taken smaller and smaller, we expect to go from the integral
conservation equation \eqref{eq:int_cons} toward the differential
continuity equation \eqref{eq:IDeq}, for which we obtained $\gd{R^{\rm (sep)}}\simeq4.5$
via Eq. \eqref{eq:Rs}. For consistency, we expect that $R^{(\rm
  CG)}(\epsilon)\to \gd{R^{\rm (sep)}}$, as the ID region is approached. The behavior of $R^{(\rm
  CG)}(\epsilon)$ is shown in Fig. \ref{fig:5}. We
observe that as more and more local interactions are left out of the
picture as the size of the \gd{coarse-graining} box becomes smaller, the direction of the coarse-grained
flux becomes more vertical, and this is consistent with Fig. \ref{fig:1},
since local colinear interactions have an enhanced horizontal
transport while the ID region has a stronger vertical
transport.  \cite{dematteis_lvov_2021} argued that
reasonably $\epsilon$ should be located between
$\frac{1}{32}$ and $\frac{1}{16}$\gd{, for what is considered ``scale-separated'' to be approximated by the Induced Diffusion approximation with an error not larger than $5-10\%$ (see Appendix B for supporting evidence)}. Notice that in Fig. \ref{fig:5} the value of $R^{(\rm CG)}$ tends exactly to the constant given by $\gd{R^{\rm (sep)}}\simeq 4.5$, and it does so for values of $\epsilon$ roughly below the chosen threshold $\epsilon=1/16$, which is thus confirmed to be about the largest value for which the ID approximation can hold. For $\epsilon < 1/16$, the diffusion coefficients scale with $\epsilon$ according to Eq. \eqref{eq:coeff_diff}, and their ratio is independent of $\epsilon$.

\subsection{Closure for the ID energy flux: rotating case}\label{sec:IDb}

So far, we have considered the non-rotating limit of the internal wave
kinetic equation \eqref{eq:kineq}. In the presence of background
rotation $f\neq 0$, scale invariance is lost and the picture is more
complex, with supplementary terms in the matrix element and a
nontrivial deformation of the resonant manifold. Since $f$ represents
the lowest internal wave frequency, having $f\neq0$ has most impact on
the 3-wave interactions involving a low frequency,  $\omega_1\sim f$. Thus, in first approximation one can assume
that the presence of background rotation affects mostly the scale
separated triads, while only marginally changing the contribution from
local triads whose three frequencies are abundantly larger than
$f$. Therefore, here we focus on the scale-separated interactions in
the rotating case, where the ID equation \eqref{eq:IDeq} represents
again the leading process. We follow a well-known derivation
(\cite{mccomas1981time,M86,polzin2017oceanic}\gd{ (section 4.6 therein)}) exploiting the
approximation of the near-inertial frequency by $f$, by which one
obtains a modified version of~\eqref{eq:diff} that reads:
\begin{equation}\label{eq:diffrot}
\begin{aligned}
    &a_{kk}=d_{kk} k^{7-a}m^{-b}\,,\quad a_{km}=d_{km} k^{6-a}m^{1-b}\,,\\ &\qquad\qquad a_{mm}=d_{mm} k^{5-a}m^{2-b}\,,
    \end{aligned}
\end{equation}
where the $d_{ij}$'s are constants. {\it E.g.}, for GM76 ($a=4,b=0$)
this yields the familiar scaling for the vertical-vertical diffusion
coefficient: $a_{mm}\propto k m^2$.

In analogy with the derivation in Sec. 4\ref{sec:IDa}, now we use \eqref{eq:diffrot} in Eq. \eqref{eq:IDeq}, again for a 2D action spectrum $n(k,m)=4\pi A k^{-a+1}m^{-b}$,  and we obtain:
\begin{equation}
\begin{aligned}
    \frac{\partial n(k,m)}{\partial t} &= -4\pi A\frac{\partial}{\partial k}\left[\left( a c_{kk} + b c_{km}\right)k^{7-2a}m^{-2b} \right]\\
    & \;\;\;-4\pi A \frac{\partial}{\partial m}\left[\left( a c_{km} + b c_{mm}\right)k^{6-2a}m^{1-2b} \right]\\
    &= 4\pi A\left[(2a-7)\left( a c_{kk} + b c_{km}\right)\right.\\
    &\qquad\;\;\; \left.+ (2b-1) \left( a c_{km} + b c_{mm}\right)\right]k^{6-2a}m^{-2b}\,.
\end{aligned}
\end{equation}
Now, at the steady state the r.h.s. must vanish for all $k$ and $m$, implying
\begin{equation}\label{eq:statrot}
    \frac{a c_{kk} + b c_{km}}{a c_{km} + b c_{mm}} = \frac{1-2b}{2a-7}\,.
\end{equation}
Use of \eqref{eq:IDeq} and \eqref{eq:diffrot} yields the stationary energy flux
\begin{equation}\label{eq:fluxenrot}
\begin{aligned}
	\bJ^{(e)}(k,m) & = 4\pi \gamma A \left( (a c_{kk}+b c_{km}) k^{8-2a} m^{-1-2b},\right.\\
	&\qquad\qquad\left.(a c_{km}+b c_{mm}) k^{7-2a} m^{-2b}\right)\,,\\
&=D_0 \left((1-2b)k^{8-2a} m^{-1-2b},\right.\\
&\qquad\;\; \left.(2a-7) k^{7-2a} m^{-2b} \right)\,,\quad D_0>0\,,
\end{aligned}
\end{equation}
where the last line is true if the solution is a stationary state
($I(a,b)=0$), ensuring the validity of the condition
\eqref{eq:statrot}.
\begin{figure}
\begin{centering}
\includegraphics[width=\linewidth]{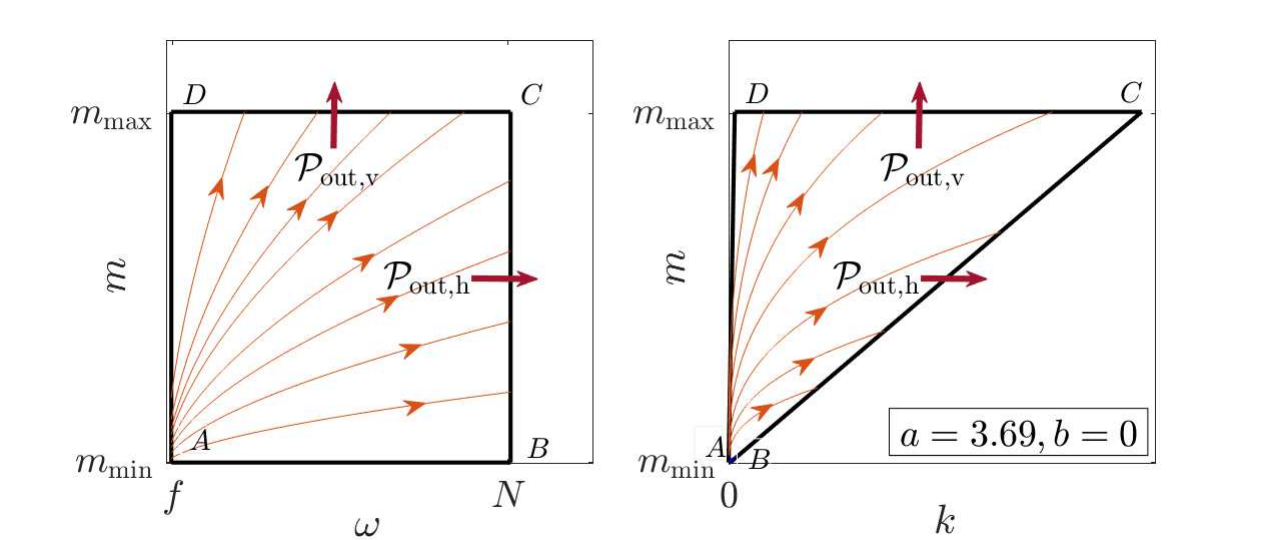}
\includegraphics[width=\linewidth]{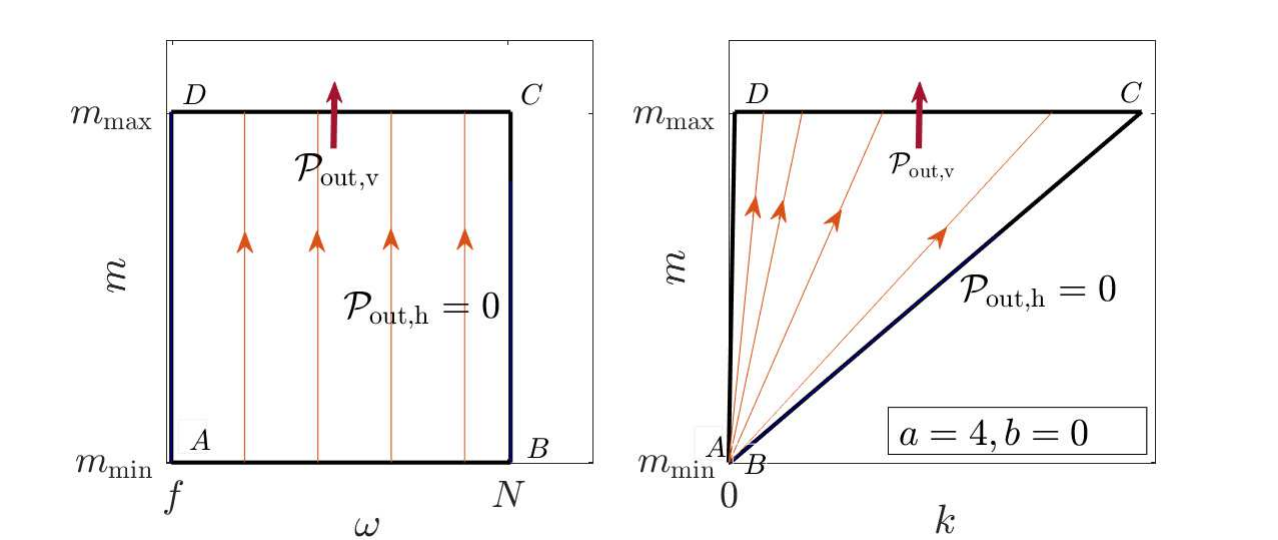}
  \caption{Direction of the ID part of the energy flux in the rotating approximation \eqref{eq:fluxenrot}. Upper panels are for the steady state of the kinetic equation; bottom panels are for the GM76 solution. As far as the downscale direction of the flux is concerned, both in vertical wavenumber {\it and} frequency, qualitative agreement with Fig. \ref{fig:4} is assessed.}
 \label{fig:6}
\end{centering}
\end{figure}

In Fig. \ref{fig:6} we show the streamlines of the \gd{rotating} ID flux \eqref{eq:fluxenrot}, for the $a=3.69,b=0$ solution (upper panels) and for the GM76 high-wavenumbers limit $a=4,b=0$ (bottom panels).  In this rotating case we proceed only as far as the dimensional analysis in Sec. 4\ref{sec:IDa}.  In the non-rotating case we have an exact \gd{power-law} solution that allows us to define a cut in the spectral domain and enables estimates of the diffusivity tensor leading to (\ref{eq:CGpowA}) and (\ref{eq:CGpowB}).  In the rotating case, $a_{mm}$ is relatively insensitive to $\epsilon$ if the cut lies, for example, at frequencies greater than $2f$, whereas $a_{ kk}$ is quite sensitive.  The absence of an exact solution in the rotating case limits greater precision.  On the other hand, we expect this result to at least provide some qualitative guidance to our intuition, \gd{indicating that a comprehensive approach to the kinetic equation with rotation (subject of current investigation) is not likely to modify sensibly the results of the present paper}. It is important to notice that \gd{the rotating approximation above confirms the downscale direction of the ID flux}, for spectra in the range between the stationary solution of the kinetic equation and GM76. In particular, we notice how the purely vertical character of the ID transport for the GM76 solution is predicted both by \eqref{eq:fluxen} and \eqref{eq:fluxenrot} (central panels of Fig. \ref{fig:4} and bottom panels of Fig. \ref{fig:6}).





\section{Summary and Discussion}\label{sec:TheEnd}

The Oceanic Ultraviolet Catastrophe originates in a first principles
asymptotic analysis of the internal wave kinetic equation
\eqref{eq:kineq} that results in the Fokker-Plank, or generalized
diffusion, equation \eqref{eq:IDeq}.  This wave-action balance
characterizes the scale-separated limit with high frequency
internal waves refracting in the vertical shear of near-inertial
waves.  As summarized in \cite{M86}, this balance leads to predictions that are at odds with observational knowledge of the oceanic internal wavefield, its sources and sinks. The analysis in this paper prioritizes the unique power-law
stationary solution $k^{-a}m^{-b} $ of the wave kinetic equation. In the 2D power-law
space $a-b$ this solution ($a\simeq3.69, b=0$) is not far from the
GM76 high-wavenumber scaling ($a=4,b=0$). Moreover, this solution is
mathematically well defined, with a collision integral (r.h.s. of the
wave kinetic equation) that is convergent, in exact balance, and
accessible to direct numerical evaluation. This exact solution has distinct contributions from both extreme scale separated interactions and interactions that are quasi-colinear in horizontal wavenumber having a more `local' character.

In the diffusive (i.e. extreme scale separated) paradigm, a further assumption that the diffusion is
dominated by the vertical-vertical coefficient leads to the onset
of no-flux solutions for $b=0$. These `no-flux' solutions include
the stationary solution of the kinetic equation and also the GM76
spectrum. Considering the fundamental use of GM76, and generalizations
thereof, to build an understanding of the observed energy fluxes
through scales, this no-flux prediction is odd enough -- representing
the first point of the Oceanic Ultraviolet Catastrophe.  Here, we have shown
that the vertical-vertical diffusive representation is an uncontrolled approximation.  Thus, before regarding these $b=0$ spectra as no-flux solutions one has to consider the other elements of the diffusion tensor. If this is
done, the flux due to Induced Diffusion turns out to be finite and
different from zero. This was shown in detail in section
\ref{sec:ID}.

In section \ref{sec:ID} we worked out a closure for the Fokker-Planck
equation based on dimensional consistency and on stationarity. This
closure provides the pointwise direction of the diffusive part of the energy flux in Fourier
space. For the stationary spectrum, the flux is downscale in both
frequency and vertical wavenumber. In particular, this is consistent
with a main source of energy localized at large vertical scales and
low frequencies. We recall that the vertical-vertical diffusion
approximation would predict energy to flow from high to low
frequencies, requiring a main energy source at high frequencies that
is believed not to be met in the oceans. Thus, the solution to the apparent paradox is once again due to the previously neglected coefficients of the diffusion tensor. Moreover, ID vertical
transport, due to the off-diagonal element of the diffusion tensor,
exceeds ID horizontal transport by a half order of
magnitude. This reveals a previously unnoticed
important role of off-diagonal diffusion in the Fokker-Planck
equation. This completes what we put forward as the solution to the
Oceanic Ultraviolet Catastrophe, but it is not the end of the story.

We have provided evidence that the reduction of the internal wave
kinetic theory to the Fokker-Planck equation, which relies on the
prominent role of the Induced Diffusion process, leaves important
contributions without extreme scale separation out of the picture. In
section \ref{sec:loc-extr} all interactions were considered. We showed
that the energy transfers can be successfully decomposed into a
`local' part and a `scale-separated' part. Independent
considerations lead to a quite distinct, non-arbitrary delimitation of
the two regions. Using the paradigm developed by \cite{dematteis_lvov_2021},
we can compute the energy fluxes at the steady state directly from
the full collision integral. All transfers, vertical and horizontal,
local and scale-separated, are directed downscale. The scale separated
part, dominated by ID, is effectively described by the
Fokker-Planck equation in section \ref{sec:ID} and gives a mainly
vertical energy flux. The local part, by far the largest contribution to the total flux, is dominated by interactions that
have near-colinear horizontal wavenumbers, as shown in
Fig. \ref{fig:3}, and has stronger horizontal transfers compared to ID (Fig. \ref{fig:5}). This
represents a novel simplified framework in which to cast local
interactions, whose effects have been shown to be far from negligible.

Despite having used a non-rotating framework throughout the manuscript, in Section 4\ref{sec:IDb} we have argued that the presence of background rotation is expected to affect mostly the contribution from scale-separated interactions. We have therefore used \gd{a well-known} approximation in the ID regime for $f\neq0$, approximating near-inertial frequencies exactly with $f$. This allowed us to obtain an alternate closure for the ID flux direction which, although nonrigorous, takes into account the background rotation. Importantly, this closure in the rotating case shares with the non-rotating case the same qualitative behavior: the direction is downscale both in vertical wavenumber and frequency, and in the GM76 case it becomes purely vertical. \gd{Independent results from \cite{regional} (their figure 38) indicate that the ``scale-separated'' low-frequency contributions play a marginal role in the overall balance, in the presence of background rotation, for a vertically homogeneous action spectrum ($b=0$). The balance appears to be mainly determined by interactions that are ``local'' in character. Both this fact and the result of Section 4\ref{sec:IDb} indicate that the non-rotating approximation of the matrix elements, is a relatively controlled approximation. Finally, one should not disregard the important benefits of the $f=0$ assumption to the rigor of the analysis. In the wave turbulence theory, when $\dot n_\bp=0$ (stationary solution of the wave kinetic equation) is associated with a Kolmogorov-Zakharov cascade, the latter is synonymous with a scale-invariant solution~\citep{ZLF}, motivating the assumption of a power law. The reason to use a non-rotating solution is that a stationary state can be defined~\citep{iwthLPTN}, which is far from obvious with the introduction of rotation.}

The close quantitative agreement of the first-principles energy fluxes, Eq. \eqref{eq:Pdematteis}, and the phenomenological finescale parameterization, Eq. \eqref{eq:Pfinescale}, deserves some last comments. The interpretation of the power dissipated horizontally is unclear. First, the boundary at $m=m_{\min}$ (refer to Fig. \ref{fig:0}) lacks a consistent major source of energy at high frequency. Therefore, in absence of a source, the upper-left corner of the box (inertial range) may not be filled with energy and as a consequence the contribution \gd{$\mP_{\out,\h}$} may be (at least in part) not realized in practice. Second, what happens at the boundary at $\omega=N$ is likely not accurately captured by the formalism in the present manuscript, as we recall that the hydrostatic approximation breaks down for $\omega\sim N$. Necessarily, a deeper understanding of this range of scales will be possible only departing from the hydrostatic approximation, but this is beyond the theoretical framework currently available. \gd{Third, although the validity of the weak nonlinearity assumption has been shown to hold for most of the inertial range (refer to the box in figure~\ref{fig:0}), it was also noticed that approaching the boundary $\omega=N$ the nonlinear time becomes of the same order of magnitude as the linear time and the weakly nonlinear resonant picture may break down (\cite{LvovNearRes, eden2019numerical}). This observation echoes the early warning by \cite{holloway1980oceanic}. On the one hand, our analysis concerns an exact stationary state for which, unlike for the nonstationary GM76 state, the ratio between linear and nonlinear time (also known as normalized Boltzmann rate) is vanishing throughout the whole $\omega-m$ space. On the other hand, our analysis is not strictly tied to the choice of the edges at $\omega=N$ or $m=m_\max$, and if a different choice is made for the integration edges in Eq.~\eqref{eq:transfer}, the modification propagates straightforwardly to Eq.~\eqref{eq:powtot}. E.g., if we move the upper edge in figure~\ref{fig:0} to $\omega=N/\sqrt{2}$ in order to avoid the above objections altogether (both to hydrostatic balance and weak nonlinearity), it is easy to see that $\mP_{\out,\h}$ increases of about $14\%$ and $\mP_{\out,\v}$ reduces of about $3.7\%$, i.e.	quite marginally. As a whole, this indicates that the breakdown of both the hydrostatic balance and the weak nonlinearity assumptions approaching $\omega=N$ should not hinder the quantitative evaluations of the current manuscript. A thorough treatment of the dependence on boundary effects, and a detailed study of the normalized Boltzmann rates is the subject of current research and is beyond the scope of the present manuscript.}

On the contrary, PSI provides a fundamental physical decay mechanism \citep{mackinnon2005subtropical,sun2013subharmonic,mackinnon2013parametric,olbers2020psi} so that the boundary $\omega=f$ can act as an energy source also at $m\gg m_{\min}$, and `fill' the lower-right corner of the inertial range. Moreover, both wave breaking and shear instability for large $m$ provide a natural pathway for the power $\mP_{\out,\v}$ to be driven towards the scales of 3D turbulence. So, the contribution $\mP_{\rm out, v} \simeq  5.2 \times 10^{-9} \, \text{\rm W  kg}^{-1}\,$ (we recall that $\mathcal{P}_{{\rm finescale}} \rightarrow 5.9\times 10^{-9} {\rm W/kg})$ appears to be better justified from different points of view, and to fully fit in the finescale-parameterization paradigm \citep{polzin2014finescale}. Concerning the dependence on the main physical parameters, we recall that $\mathcal{P}_{{\rm finescale}}$ scales as $ f N^2 \hat E^2$, Eq.~\eqref{eq:FinescaleParameterization}. Since this scaling is derived for the GM76 spectrum ($a=4,b=0$), we can consider the scaling of $\mP_{\out,\v}$ for $a=4$ ({\it i.e.} $\nu=2a-7=1$), which gives exactly $f N^2 E^2$ (we recall that $\hat E$, besides being a metric for the shear scale length, is also a measure of the spectral level, in units
of the GM76 standard spectral level). This exact scaling agreement establishes a deeper connection between the phenomenological and the first-principles estimates.

\gd{The accuracy of the kinetic equation for the extreme scale separated interactions may be affected by Doppler shifting and
modification of the Galilean invariance \citep{kraichnan1959structure,kraichnan1965lagrangian}. These effects are encapsulated in the resonant
bandwidth being proportional to the Doppler shift, as reported in \cite{polzin2017oceanic}. This question is left for future
research.}

\gd{Our efforts implement the theoretical program suggested by Ferris
 \cite{webster1969turbulence}, where ``{\it due to the lack of an adequate theoretical framework for describing turbulence in a stratified fluid}'' homogeneous three-dimensional turbulence estimates were employed; with today's internal wave turbulence, over five decades later, we are able to fully exploit the potential of the theory that the seminal contribution was advocating for}.

In summary, we have established the presence of extreme scale separated and local interactions in the internal wave kinetic equation and have shown that
    \begin{itemize}
    \item Concerning scale-separated interactions, the Fokker Planck equation and the Induced Diffusion picture of
      \cite{McComas1977} provides a remarkably good characterization
      of the dominant contributions to the internal wave scattering.
    \item The reduction of the diffusion tensor to a single vertical
      component necessitates a high frequency source of energy and
      dominance of inverse energy cascade. Both of these effects are
      non-intuitive and lack experimental evidence.
    \item Taking into account the full  diffusion
      tensor leads to direct energy cascade consistent with our
      understanding of the internal wave scattering.
      \item The vertically homogeneous $b=0$ wave action was termed the
      `no-flux' solution by \cite{mccomas1981time} due to the
      properties of the Fokker-Planck equation. Taking into account
      the complete diffusion tensor in both vertical and horizontal direction
      does create non-zero vertical and horizontal energy fluxes. 
    \item Induced diffusion, however, does not capture all the
      processes that contribute to the direct energy cascade.  Local
      interactions, in particular those with near-colinear
      horizontal wavenumbers, actually provide the majority of the total energy
      transfers.
    \item Considering the energy balances in a finite size box allows us to
      quantify numerically the magnitude and
      direction of the direct energy cascade. Taking the limit of small
      box size reproduces the Induced Diffusion limit.
    \item Numerical calculation of the total direct energy cascade
      generated by the internal wave kinetic equation leads to a (first-principles) formula which is
      remarkably close to the celebrated (phenomenological) finescale parameterization for
      the energy flux \citep{gregg1989scaling,henyey1991scaling,polzin1995finescale}.
      \end{itemize}

\acknowledgments
    {The authors gratefully acknowledge support from
      the } ONR grant N00014-17-1-2852. YL gratefully acknowledges support from NSF DMS award 2009418.


%
%

\noindent{\it Conflicts of interest statement. } The authors declare no conflicts of interest.

\datastatement
No data were created for this effort.  


%
\appendix

\appendix[A]

\appendixtitle{Matrix elements and resonant manifold}

The two delta functions in Eq.~\eqref{eq:kineq} can be integrated out analytically, obtaining
{\begin{equation}\label{eq:A3}
\begin{aligned}
	&\partial_t (kn_\bp) = \int_0^\infty dk_{1}dk_2 \; {\mJ}(k,k_1,k_2,m)\,,\\
&{\mJ}(k,k_1,k_2,m)=\left(R^0_{12} f^0_{12} - R^1_{02}f^1_{02} - R^2_{01}
        f^2_{01}\right)\,,\\
& R^0_{12} =8\pi k k_1 k_2
        |V^0_{12}|^2/\left(|{g^0_{12}}'|\Delta_{012}\right)\,.
\end{aligned}
\end{equation}}
Here $ f^0_{12} = n_1n_2 - n_\bp (n_1 + n_2)\,$ is the spectrum-dependent term of the equation,
and the area of the triangle of sides $k,k_1,k_2$, coming from
integration over angles under the assumption of isotropy, is given by Heron's formula
\begin{equation}
	\Delta_{012} = \frac12 \sqrt{2(k^2 k_1^2 + k^2 k_2^2 + k_1^2 k_2^2)-k^4-k_1^4-k_2^4}\,.\label{Delta}
\end{equation}
The expression of the so-called matrix elements $V^\bp_{\bp_1 \bp_2}$ in the scale-invariant regime  reads (~\cite{iwthLPTN})
\begin{equation}
\begin{aligned}
	&V^0_{\bp_1 \bp_2} = \sqrt{kk_1k_2}\left(
  \frac{k^2+k_1^2-k_2^2}{2kk_1}\sqrt{\left|\frac{m_2^\star}{mm_1^\star}\right|}\right.
  \\
&\left.+
  \frac{k^2+k_2^2-k_1^2}{2kk_2}\sqrt{\left|\frac{m_1^\star}{mm_2^\star}\right|}
  + \frac{k^2-k_1^2-k_2^2}{2k_1k_2}\sqrt{\left|\frac{m}{m_1^\star
      m_2^\star}\right|} \right) \,,\\
	\end{aligned}
\end{equation}
and, moreover, we have
\begin{equation}
	{g^0_{12}}' = \frac{\sign (m_1^\star)\; k_1}{(m_1^\star)^2}
        - \frac{\sign (m_2^\star)\; k_2}{(m_2^\star)^2} \,,
\end{equation}
where $m_1^\star, m_2^\star$ are given by the solution of the
resonance conditions, i.e. the joint conservation of momentum and
energy in each triadic resonant interaction. Thus, in the four-dimensional
space spanned by $k_1$, $k_2$, $m_1$, $m_2$, the problem is
restricted to the {\it resonant manifold}, parametrized by two
independent variables $k_1$ and $k_2$ as summarized in Table A0.
%
\begin{table*}
\begin{center}
\begin{tabular}{ c |c |c }
Label $\;$&$\;$ Resonance condition$\;$ &$\;$ Solutions $\;$ \\ \hline
 $(\rm{Ia}),(\rm{Ib})$ & $\left\{\begin{array}{ll}
	\bp = \bp_1 + \bp_2\,\\
	\frac{k}{|m|} = \frac{k_1}{|m_1|} + \frac{k_2}{|m-m_1|}
\end{array}\right.$ & $\left\{\begin{aligned}
	&m_1^\star = \frac{m}{2k}\left[ k \pm k_1 \pm k_2 \pm \sqrt{(k\pm k_1 \pm k_2)^2 \mp 4 kk_1} \right]\\
	&m_2^\star = m - m_1^\star
\end{aligned}\right.$ \\  \hline
 $(\rm{IIa}),(\rm{IIb})$ & $\left\{\begin{array}{ll}
	\bp_1 = \bp + \bp_2\,\\
	\frac{k_1}{|m_1|} = \frac{k}{|m|} + \frac{k_2}{|m_1-m|}
\end{array}\right.$   & $ \left\{\begin{aligned} &m_2^\star
        = -\frac{m}{2k}\left[ k \mp k_1 - k_2 + \sqrt{(k\mp k_1 - k_2)^2 +
            4 kk_2} \right]\\ &m_1^\star = m + m_2^\star
\end{aligned}\right.\ $\\\hline
 $(\rm{IIIa}),(\rm{IIIb})$ & $\left\{\begin{array}{ll}
	\bp_2 = \bp + \bp_1\,\\
	\frac{k_2}{|m_2|} = \frac{k}{|m|} + \frac{k_1}{|m_2-m|}
\end{array}\right.$  &  $\left\{\begin{aligned}
	&m_1^\star = -\frac{m}{2k}\left[ k - k_1 \mp k_2 + \sqrt{(k-k_1 \mp k_2)^2 + 4 kk_1} \right]\\
	&m_2^\star = m + m_1^\star
\end{aligned}\right. $
\end{tabular}
\appendcaption{A1}{The six independent solutions to the resonance conditions, defining the {\it resonant manifold} in the space spanned by the two free variables $k_1,k_2$. }\label{tab:A1}
\end{center}
\end{table*}
Note the symmetries of the resonant manifold: the solution $(\rm Ia)$
is obtained from solution $(\rm Ib)$ through permutation of the
indices $1\leftrightarrow 2$. We also notice that solutions $(\rm
IIa)$, $(\rm IIb)$ reduce to solutions $(\rm IIIa)$, $(\rm IIIb)$,
respectively, under permutation of the indices $1\leftrightarrow2$.

The collision integral of Eq.~\eqref{eq:A3} is integrated over the so-called ``kinematic box'', represented in Fig.~\eqref{fig:2}. 

\appendix[B]
\appendixtitle{Region of validity of the ID asymptotics}

In the IR region (Fig.~\ref{fig:2}) the two resonant Induced Diffusion branches $({\rm Ia})$ and $({\rm IIa})$ (refer to Table 1) dominate over the others and we adopt the following change of variables
\begin{equation}
	k_1=k(1+y),\qquad k_2= kx\,,
\end{equation}
with $0<x<\epsilon, -x<y<x$, that allows us to use the following Taylor expansions for the conditions $({\rm Ia})$ and $({\rm IIa})$, respectively,
\begin{equation}\label{eq:A11}
	m_1^\star \simeq m(1+\sqrt{x}+\frac12 (x+y)),\qquad m_2^\star \simeq -m(\sqrt{x}+\frac12 (x+y))\,,
\end{equation}
\begin{equation}\label{eq:A12}
	m_1^\star \simeq m(1-\sqrt{x}+\frac12 (x+y)),\qquad m_2^\star \simeq -m(\sqrt{x}-\frac12 (x+y))\,,
\end{equation}
using the fact that $x=O(\epsilon), y=O(\epsilon)$. In the rest of the section, we use the short-hand notation $m_1^\star=m(1+\eta),m_2^\star=-m\eta$, where $\eta=\pm\sqrt{x}+\frac12(x+y)=O(\sqrt{\epsilon})$, for~\eqref{eq:A11} and~\eqref{eq:A12}, respectively. With the asymptotics of Eqs.~\eqref{eq:A11}-\eqref{eq:A12}, neglecting the lower order term $R^2_{01} f^2_{01}$ and Taylor expading the spectrum-dependent terms in the collision integral around the point $(x,y)=(0,0)$, we obtain
\begin{equation}\label{eq:A11a}
\begin{aligned}
	&f^0_{12}\simeq n(kx,-m\eta) \left(ky \frac{\partial n}{\partial k}+m\eta \frac{\partial n}{\partial m}\right)\,,\\
& f^1_{02}\simeq -n(kx,-m\eta) \left(ky \frac{\partial n}{\partial k}+m\eta \frac{\partial n}{\partial m}\right)\,,
\end{aligned}
\end{equation}
which implies
\begin{equation}\label{eq:A12a}
\begin{aligned}
	&{\mJ}(k,k_1,k_2,m)\\&\quad\simeq \left(R^0_{12} + R^1_{02}\right) \; n(kx,-m\eta) \left(ky \frac{\partial n}{\partial k}+m\eta \frac{\partial n}{\partial m}\right)\,.
\end{aligned}
\end{equation}
The leading order expressions of $R^0_{12}$ and $R^1_{02}$ (on which the matrix elements depend) are given by
\begin{equation}\label{eq:A27a}
	R^0_{12} \simeq 8\pi\left[ \frac{2k^3my^2}{\sqrt x \sqrt{x^2-y^2}} - \frac{k^3 m (2x^2 y -2x y^2 - y^3/4)}{ x \sqrt{x^2-y^2}}\right]\,,
\end{equation}
\begin{equation}\label{eq:A27b}
	R^1_{02} \simeq  8\pi\left[\frac{2k^3my^2}{\sqrt x \sqrt{x^2-y^2}} + \frac{k^3 m (2x^2 y -2x y^2 - y^3/4)}{ x \sqrt{x^2-y^2}}\right]\,.
\end{equation}
Some algebra and one further Taylor expansion allow us to quantify the diffusion coefficients at the stationary state for Eq.~\eqref{eq:IDeq}, with result given in Eq.~\eqref{eq:coeff_diff}.
\begin{figure*}
\begin{centering}
\includegraphics[width=1\linewidth]{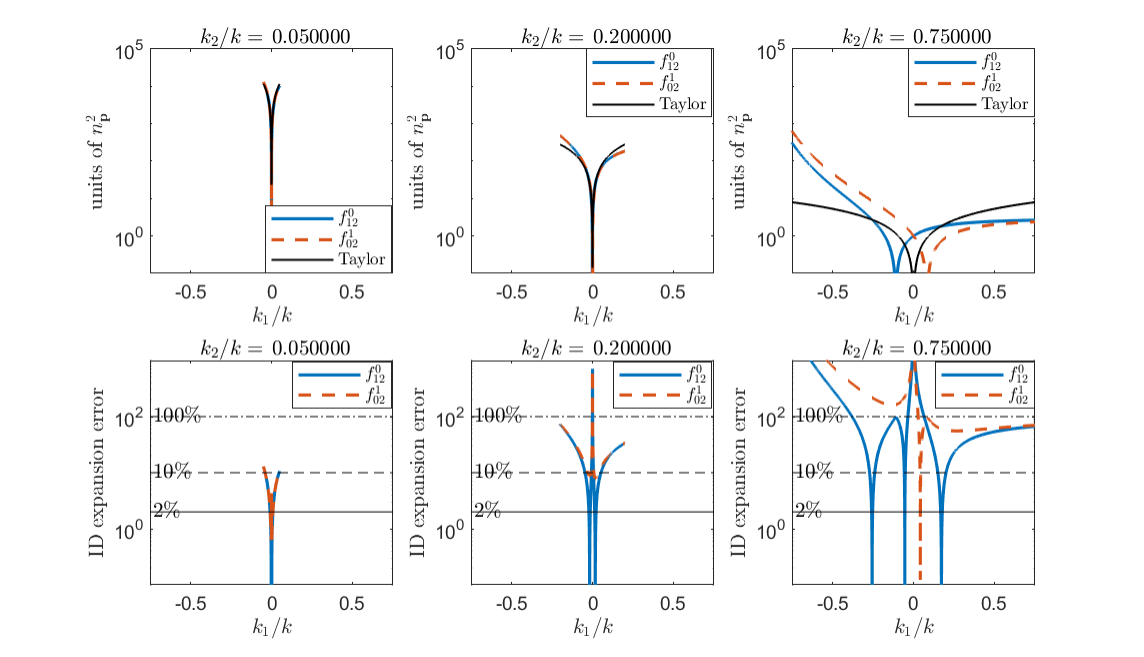}
  \appendcaption{A1}{Top: comparison of the numerically computed functions $f^0_{12}$ and $f^1_{02}$ and their respective leading orders as given in Eq.~\eqref{eq:A11a}, as a function of $y=k_1/k-1$, for three different values of $x=k_2/k$. Bottom: relative errors of the leading order estimates in the upper panels.} \label{fig:A4}
\end{centering}
\end{figure*}
In Fig.~\ref{fig:A4} we propose a simple test to establish the region of validity of the approximation~\eqref{eq:A11a}, for the solution $(a,b)=(3.69,0)$. The quantities $f^0_{12}$ and $f^1_{02}$ are computed numerically and compared with their leading order approximation given in Eq.~\eqref{eq:A11a}, for three different values of $x=k_2/k$, as a function of $y=k_1/k-1$. To visualize this in the kinematic box one can look at Fig.~\ref{fig:2}, and move horizontally on a section at fixed $x$. The boundaries at $y=\pm x$ are the locations where the plotted functions are largest. The error of the estimate is about $10\%$ at the boundaries of the section with $x=1/20$. At the boundaries of the section with $x=1/5$ the error is in the range $30-80 \%$, and the error is out of control (above $100\%$) when $x=3/4$. This shows that a diffusion closure is not possible for interactions in the kinematic box above $k_2/k\simeq0.1$, i.e. outside the IR region of Fig.~\ref{fig:2}. As a consequence, it is not possible to extend the integration region of the integrals defining $a_{kk}$ and $a_{mk}$ to larger values of $\epsilon$, since for $\epsilon >0.1 $ the diffusive character of the interaction is gradually lost. We remark that this fact has been known since the original derivation of \cite{McComas1977}, where in the definition of the diffusion coefficients the small-wavenumber part of the spectrum $B(\bp)$ is present, and not the full spectrum $n(\bp)$. Previously in the paper, $B(\bp)$ is defined as the restriction of $n(\bp)$ for ``small wavenumbers''. Our results illustrate that $B(\bp)$ is the restriction of $n(\bp)$ to the IR region. The rest of the contributions are ``local'' interactions as defined in Eq.~\eqref{eq:kineq_decomp}. The choice of $\epsilon=1/16$ to demark the separation between the two regions named ``local'' and ``scale-separated'' corresponds to an error of the approximation~\eqref{eq:A11a} around $10\%$, meaning that our ID approximation to the scale-separated contribution, as used in this manuscript, is ``controlled'' by an error of at most $10\%$.

%




%
%
%
\bibliographystyle{ametsoc2014}
\bibliography{references}

%

%

\end{document}